\documentclass[]{jfm}

\usepackage{graphicx}
\usepackage{newtxtext}
\usepackage{newtxmath}
\usepackage{natbib}
\usepackage{hyperref}
\hypersetup{
    colorlinks = true,
    urlcolor   = blue,
    citecolor  = black,
}

\newcommand{\RomanNumeralCaps}[1]
\linenumbers
\usepackage{subfig}
 \usepackage[abs]{overpic}
\usepackage{placeins}
\usepackage{tikz}
\usepackage{bm}

\title{Understanding the effect of Prandtl number on momentum and scalar mixing rates in neutral and stably stratified flows using gradient field dynamics}

\author{Andrew D. Bragg\aff{1}\corresp{\email{andrew.bragg@duke.edu}}
 \and Stephen M. de Bruyn Kops\aff{2}
 }

\affiliation{\aff{1}Department of Civil and Environmental Engineering, Duke University, Durham, NC 27708, USA
\aff{2}Department of Mechanical and Industrial Engineering, University of Massachusetts Amherst, Amherst, MA 01003, USA
}

\begin{document}

\maketitle
\begin{abstract}
Recently, direct numerical simulations (DNS) of stably stratified turbulence have shown that as the Prandtl number ($Pr$) is increased from 1 to 7, the mean turbulent potential energy dissipation rate (TPE-DR) drops dramatically, while the mean turbulent kinetic energy dissipation rate (TKE-DR) increases significantly. Through an analysis of the equations governing the fluctuating velocity and density gradients we provide a mechanistic explanation for this surprising behavior and test the predictions using DNS. We show that the mean density gradient gives rise to a mechanism that opposes the production of fluctuating density gradients, and this is connected to the emergence of ramp-cliffs. The same term appears in the velocity gradient equation but with the opposite sign, and is the contribution from buoyancy. This term is ultimately the reason why the TPE-DR reduces while the TKE-DR increases with increasing $Pr$. Our analysis also predicts that the effects of buoyancy on the smallest scales of the flow become stronger as $Pr$ is increased, and this is confirmed by our DNS data. A consequence of this is that the standard buoyancy Reynolds number does not correctly estimate the impact of buoyancy at the smallest scales when $Pr$ deviates from 1, and we derive a suitable alternative parameter. Finally, an analysis of the filtered gradient equations reveals that the mean density gradient term changes sign at sufficiently large scales, such that buoyancy acts as a source for velocity gradients at small scales, but as a sink at large scales.
\end{abstract}

\section{Introduction}

In simple fluids where molecular transport is modeled as a gradient-diffusion process, the mixing rates of quantities such as momentum, heat and species are determined by the associated molecular diffusion coefficient and the magnitude of spatial gradients of the quantity. In a turbulent flow, complex stirring motions lead to the intensification of spatial gradients of flow quantities, which in turn 
enhances the mixing rates. In this sense, the mixing rates are controlled by the stirring processes themselves. This fact is often exploited when modeling mixing rates because the wide range of dynamically relevant length and time scales in high Reynolds number turbulent flows means that the small-scale mixing often cannot be directly resolved and so it is instead modeled indirectly based on stirring rates at resolved scales. This assumption underlies the classical $k$-$\epsilon$ closure for the Reynolds Averaged Navier Stokes equations (RANS), as well as models based on a turbulent Prandtl number (we do not distinguish between heat and species and use the term Prandtl number for both).  Two-point closures for RANS and conditional moment closure are examples of approaches that do not directly couple mixing and stirring rates, but, nevertheless, the former is inferred from the latter without information about dynamics at the smallest scales where the mixing actually takes place. 
 
The motivation for the research reported here is that in stably stratified flows (subject to the Boussinesq approximation), varying the diffusion coefficient of the scalar has been observed to affect the mixing rates of not only the scalar but also of momentum.  In the very simple configuration of initially homogeneous and isotropic turbulence subjected to a stabilizing density gradient, \citet{riley23} find that not only is the dissipation rate of potential energy significantly lower at Prandtl number $Pr=7$ than at $Pr=1$, but the dissipation rate of kinetic energy is also higher at $Pr=7$.  In fact, it has been known for some time that higher $Pr$ results in slower mixing of heat in stratified flows \citep{smyth01}.  More recently, \citet{salehipour15} found that $Pr$ has a strong effect on secondary instabilities in stratified flows, and \citet{legaspi20} observed transfer of potential to kinetic energy at small scales that depends on $Pr$.   
 
An interesting feature of the homogeneous flow studied by \citet{riley23} is that the large-scale structures are not obviously affected by the changes in $Pr$ other than that they lose energy at differing rates depending on $Pr$. But if mixing rates are determined by stirring rates, then since the mixing rates were observed in \citet{riley23} to depend strongly on $Pr$, the stirring rates at some scales in the flow must also be strongly affected by $Pr$. The connection between stirring and mixing rates in stratified turbulence has been traditionally approached from the perspective of multiscale flow energetics, i.e. analyzing kinetic and potential energies using Fourier analysis. However, to understand the physical mechanism by which stirring and mixing rates in stratified turbulence are affected by $Pr$, we find it more insightful to study the problem by analyzing the equations governing velocity and scalar gradients in the flow. Production mechanisms in these equations are associated with the stirring processes that intensify flow gradients, and the magnitude of the resulting gradients determines the mixing rates. 

In the context of homogeneous, isotropic turbulence, studying turbulent flows from the perspective of velocity gradient dynamics has a long and rich history that has led to numerous insights into the physics of small-scale turbulence \citep{vieillefosse1982local,ashurst87, nomura_post_1998,chertkov1999lagrangian,tsinober_book, chevillard2006lagrangian, gulitski2007velocity, meneveau2011lagrangian,danish18,carbone_iovieno_bragg_2020,tom21}. For stratified flows where the momentum and density fields are coupled, velocity gradient dynamics would need to be studied in conjunction with those of density gradients, and very little has been done on this. Recent notable exceptions are the insightful studies of \citet{Sujovolsky19, Sujovolsky20,Marino22}. In these, simplified forms of the velocity and density gradient equations were considered in which molecular transport and the non-local pressure Hessian terms were discarded (similar in spirit to the Restricted Euler model of \citet{vieillefosse1982local}). For the resulting simplified model, invariant manifolds were discovered, and the way that phase-space trajectories move between these manifolds was shown to explain the enhanced intermittency and marginal instability that has been observed in stably stratified flows when the Froude number is within a certain range \citep{Rorai14,Feraco_2018}.

In our study we will analyze the exact (within the Boussinesq framework) forms of the coupled velocity and density gradient equations in order to understand the mechanism responsible for the strong $Pr$ dependence of mixing rates in stably stratified turbulence observed in \citet{riley23}. It will be shown that the mechanism is associated with the competition between distinct production terms in the gradient equations that are associated with either the fluctuating or mean density gradient field. The term associated with the mean density gradient actually opposes the production of fluctuating density gradients, and this is ultimately the effect responsible for the momentum mixing rate increasing and the density mixing rate decreasing as $Pr$ is increased, as observed in \citet{riley23}. Furthermore, we also study the behavior of velocity and passive scalar gradients in the context of stationary, homogeneous, isotropic turbulence with a mean scalar gradient. It will be seen that the mechanism responsible for the striking effect of $Pr$ on scalar mixing rates in stratified turbulence is in fact already present even in the case of a passive scalar. It is simply that this mechanism plays a very small role in the passive scalar case, although it could play an important role even in that case depending upon the parameter regime of the flow.

\section{Theory: gradient dynamics in neutral flows}
So that we can consider passive and active scalars using the same notation, let the scalar in all cases be density $\rho$ assuming the non-hydrostatic Boussinesq approximation, where the gravitational acceleration is zero for case of a passive scalar.  Then $\rho=\rho_r+\gamma z+\varrho$, where $\rho_r$ is the reference density, and $\varrho$ is the fluctuation about the mean density $\langle\rho\rangle=\rho_r+\gamma z$, with $\gamma$ a constant. The equations for the velocity $\bm{u}$ and density fluctuations $\varrho$ are
\begin{align}
D_t\bm{u}&=-\bm{\nabla}p+\nu\nabla^2\bm{u}-\varrho\rho_r^{-1}g\bm{e}_z+\bm{F},
\label{eq:eq1}
\\
D_t\varrho&=\nu Pr^{-1}\nabla^2\varrho-\gamma u_z,
\end{align}
where $D_t\equiv \partial_t+(\bm{u\cdot\nabla})$ is the Lagrangian derivative, $p$ is the pressure, $\nu$ is the kinematic viscosity, $\bm{e}_z$ is the unit vector in the vertical direction, $\bm{F}$ is a forcing term, and $Pr$ is the Prandtl number. It is convenient to introduce the variable $\phi\equiv \varrho/(\beta'\rho_r)$, where $\beta'\equiv\sqrt{-\gamma/\rho_r}$, which is related to the buoyancy frequency $N$ through the relation $N=\sqrt{g}\beta'$. When non-dimensionalized using time-independent, large-eddy length $L$ and velocity $U$ scales, the equations for $\bm{u}$ and $\phi$ may be written as
\begin{align}
D_t\bm{u}&=-\bm{\nabla}p+Re^{-1}\nabla^2\bm{u}- \beta Fr^{-2}\phi\bm{e}_z+\bm{F},\label{u_nond}\\
D_t\phi&=(Pr Re)^{-1}\nabla^2\phi+\beta u_z,\label{phi_nond}
\end{align}
where $Re\equiv LU/\nu$ is the Reynolds number, $Fr\equiv U/\sqrt{gL}$ is the Froude number, $\beta\equiv \beta' \sqrt{L}$, and all variables here and hereafter are in non-dimensional form (for notational simplicity we do not distinguish non-dimensional variables, e.g. by using $\widetilde{\cdot}$). $L$ and $U$ will be taken to be the horizontal integral length scale and the horizontal root-mean-square velocity, respectively, since these are suitable choices when analyzing stably stratified flows. For the non-stationary stratified flows considered later, the values of $L,U$ at the instant the flow starts to decay are used. Note that in stratified turbulence studies it is common to use a Froude number based on the buoyancy frequency $Fr_N\equiv U/LN$ which is related to the variables introduced above through $Fr_N=Fr/\beta$.

For statistically homogeneous flows (as considered in this paper), the equations governing the average kinetic energy (per unit-mass) $\langle\|\bm{u}\|^2\rangle/2$ and ``scalar energy'' $\langle\phi^2\rangle/2$ are
\begin{align}
(1/2)\partial_t\langle\|\bm{u}\|^2\rangle&=-2 Re^{-1}\langle\|\bm{S}\|^2\rangle- \beta Fr^{-2}\langle \phi u_z\rangle+\langle\bm{F}\bm{\cdot u}\rangle,\label{KE}\\
(1/2)\partial_t\langle\phi^2\rangle &=-(Pr Re)^{-1}\langle\|\bm{B}\|^2\rangle+\beta \langle\phi u_z\rangle,\label{SE}
\end{align}
where $\bm{S}\equiv (\bm{\nabla u}+\bm{\nabla u}^\top)/2$ is the strain-rate tensor, and $\bm{B}\equiv\bm{\nabla}\phi$. 

In equations \eqref{KE} and \eqref{SE}, the energy dissipation rates are $\langle\epsilon\rangle\equiv 2Re^{-1}\langle\|\bm{S}\|^2\rangle$ and $\langle\chi\rangle\equiv (PrRe)^{-1}\langle\|\bm{B}\|^2\rangle$. In the context of stratified flows, $Fr^{-2}\langle\phi^2\rangle/2$ corresponds to the mean turbulent potential energy in the flow and $Fr^{-2}\langle\chi\rangle$ is its dissipation rate. One of the key goals of this work is to understand the mechanisms controlling $\langle\epsilon\rangle$ and $\langle\chi\rangle$ and how they depend upon $Pr$. Since these dissipation rates are fundamentally related to the gradients $\bm{A}\equiv\bm{\nabla u}$ and $\bm{B}\equiv\bm{\nabla}\phi$, it is the behavior of these gradients that must be understood in order to understand the dissipation rates and their dependence on $Pr$. The equations governing the gradients are

\begin{align}
D_t\bm{A}&=-\bm{A\cdot A}-\bm{\nabla\nabla}p+Re^{-1}\nabla^2\bm{A}-\beta Fr^{-2}\bm{B}\bm{e}_z+\bm{\nabla F},\label{Aeq}\\
D_t\bm{B}&=-\bm{A}^\top\bm{\cdot B}+(Pr Re)^{-1}\nabla^2\bm{B}+\beta\bm{A}^\top \bm{\cdot e}_z,\label{Beq}
\end{align}
and the role of each of the terms in these equations will be discussed in the analysis that follows.

We will begin by considering the dynamics of neutrally buoyant flows $Fr=\infty$ for which the scalar is passive, since it will be shown that some of the key properties of a passive scalar driven by a mean gradient play an important role in the behavior of stratified flows. For the passive scalar case it will be assumed that the forcing $\bm{F}$ generates a statistically stationary, isotropic turbulent flow. We will also consider the case where the scalars are introduced to the steady flow with $\bm{B}(0)=\bm{0}$ since this is the situation that will be considered later in the DNS of decaying stratified turbulence, and we want to understand how $\bm{B}$ evolves from its initial state to its stationary behavior. Note that for the passive scalar case the statistics of $\bm{B}$ change trivially under the transformation $\gamma\to-\gamma$, and so for consistency with the stably stratified case we only consider $\gamma<0$ in the analysis that follows such that $\beta\in\mathbb{R}^+$. 

\subsection{Impact of the Batchelor regime}\label{IBR}

When $Pr\neq 1$ there is a difference between the smallest scales of the momentum and scalar fields. While the smallest scale (in a mean-field sense) of the momentum field is the Kolmogorov scale $\eta$, the smallest scale of the scalar field is the Batchelor scale $\eta_B=Pr^{-1/2}\eta$ when $Pr\geq 1$ \citep{batchelor59a}, while for $Pr<1$ it is the Obukhov-Corrsin scale $\eta_{OC}=Pr^{-3/4}\eta$ \citep{corrsin51,obukhov49}. When $Pr\gg 1$, there is a separation of scales $\eta\gg\eta_B$ corresponding to the so-called ``viscous-convective range'' in which the effects of viscosity are important, but the effects of molecular diffusion on the scalar field are not. In terms of equation \eqref{Beq}, the significance of this is that for the term $-\bm{A}^\top\bm{\cdot B}$, which describes how the fluctuating velocity gradients amplify (or suppress) the fluctuating scalar gradients, $\bm{A}$ and $\bm{B}$ may exhibit fluctuations at different scales in the flow. When $Pr\gg1$, $\bm{B}$ will exhibit fluctuations on a much finer scale than $\bm{A}$, on average, and this ``de-localization'' between the scale at which $\bm{A}$ and $\bm{B}$ fluctuate impacts the behavior of $-\bm{A}^\top\bm{\cdot B}$. This de-localization effect was previously considered in \citet{nazarenko_laval_2000} for passive scalars in two-dimensional turbulence using Fourier analysis, rather than the gradient fields as discussed here. 

The de-localization effect that arises in the viscous-convective regime can impact the $Pr$ dependence of $\langle\chi\rangle$. In \citet{donzis05} a model for $\langle\chi\rangle$ was presented that captures this effect phenomenologically. In particular, for the case of $Pr\geq 1$, the scalar spectrum in the inertial-convective range (where the effects of $\nu$ and $Pr$ are both assumed to be unimportant) was modeled using a Obhukov-Corrsin spectrum  \citep{corrsin51,obukhov49}, and that in the viscous-convective range was modeled using a Batchelor spectrum, leading to (here $\langle\chi\rangle$ is dimensional)
\begin{align}
\frac{L}{U}\frac{\langle\chi\rangle}{\langle\phi^2\rangle}\sim\frac{1}{c_1\Big(f^{2/3}-c_3 Re_\lambda^{-1} \Big)+c_2Re_\lambda^{-1} \ln Pr},\label{Donzis_model}
\end{align}
where $Re_\lambda$ is the Taylor Reynolds number, $f\equiv A(1+\sqrt{1+(B/Re_\lambda)^2})$, and $A\approx 0.2, B\approx 92, c_1\approx 0.6, c_2\approx (5/3)\sqrt{15}, c_3\approx \sqrt{15}$. These values were determined by fitting the model to the DNS data (since the assumed spectrums involve unknown coefficients), except for the factors involving $\sqrt{15}$ which arise due to isotropy of the flow.

The $\ln Pr$ dependence in \eqref{Donzis_model} arises from the contribution due to the Batchelor spectrum for the viscous-convective range. This model predicts that for finite $Pr$, $\lim_{Re_\lambda\to \infty}[L\langle\chi\rangle/(U\langle\phi^2\rangle)]\sim 1/(c_1 4^{1/3} A^{2/3})$, i.e. a constant reflecting anomalous behavior in this limit. However, for finite $Re_\lambda$ it predicts $\lim_{Pr\to \infty}[L\langle\chi\rangle/(U\langle\phi^2\rangle)]\sim Re_\lambda/(c_2\ln Pr)$, i.e. no dissipation anomaly. This logarithmic behavior was confirmed in \citet{donzis05} at low $Re_\lambda$, and more recently in \citet{Buaria21b} at a higher Reynolds number $Re_\lambda=140$ over the range $Pr\in[1,512]$. In view of the derivation of \eqref{Donzis_model}, the interpretation is that the behavior of $L\langle\chi\rangle/(U\langle\phi^2\rangle)$ will only be anomalous when the Batchelor regime of the scalar spectrum makes a sub-leading contribution to $\langle\chi\rangle$, and the Obhukov-Corrsin regime dominates.

In addition to the model in \eqref{Donzis_model}, \citet{donzis05} also derived a model for $\langle\chi\rangle$ that applies for $Pr<1$ by integrating the Obhukov-Corrsin spectrum up to the cut-off wavenumber $k\sim 1/\eta_{OC}$. This model also predicts a $Pr$ dependence of $\langle\chi\rangle$, however, in this case it involves $Pr^{1/2}$ rather than the $\ln Pr$ factor that arises for $Pr\geq 1$. The $Pr$ dependence of $\langle\chi\rangle$ only vanishes in the regime $Pr<1$ when $Re_\lambda Pr^{1/2}$ is sufficiently large.

\subsection{Behavior of production terms and the role of ramp-cliff structures}\label{ERCS}

In addition to the de-localization effect that influences the behavior of $-\bm{A}^\top\bm{\cdot B}$ in \eqref{Beq} when $Pr\neq 1$, there is a second way in which $Pr$ can influence the stirring processes that govern the amplification of $\bm{B}$, which in turn can influence the $Pr$ dependence of $\langle\chi\rangle$. This second effect arises due to a $Pr$-dependent competition between $-\bm{A}^\top\bm{\cdot B}$ and $\beta\bm{A}^\top \bm{\cdot e}_z$ in \eqref{Beq}. This effect was not accounted for in the model of \citet{donzis05} for $\langle\chi\rangle$ because they assumed that the mean scalar gradient is unimportant for the behavior of $\langle\chi\rangle$. While we will ultimately show that for passive scalars this second effect is indeed usually unimportant, we explain it in significant detail here because it will be shown that it is in fact the main contributor to the strong $Pr$ dependence of $\langle\chi\rangle$ observed for stratified flows in \citet{riley23}. This therefore provides mechanistic insights into how scalar mixing can differ in significant ways for neutral and stratified flows.

From \eqref{Beq} we obtain
\begin{align}
\frac{1}{2}D_t\|\bm{B}\|^2&=\mathcal{P}_{B1} +\mathcal{P}_{B2} +(2 Pr Re)^{-1}\nabla^2\|\bm{B}\|^2-\mathcal{D}_{B},\label{Bsquared}
\end{align}
where $\mathcal{P}_{B1} \equiv-\bm{B\cdot}\bm{A}^\top\bm{\cdot}\bm{B}$ is the production term associated with the fluctuating scalar gradient, $\mathcal{P}_{B2} \equiv \beta\bm{B\cdot}\bm{A}^\top\bm{\cdot}\bm{e}_z$ is the production term associated with the mean scalar gradient, and $\mathcal{D}_{B} \equiv (Pr Re)^{-1}\|\bm{\nabla}\bm{B}\|^2$ is the dissipation rate of $\|\bm{B}\|^2$.

For a statistically homogeneous flow 
\begin{align}
\frac{1}{2}\partial_t\langle\|\bm{B}\|^2\rangle=\langle \mathcal{P}_{B1}\rangle+\langle \mathcal{P}_{B2}\rangle-\langle \mathcal{D}_{B}\rangle.\label{BBeq_stationary}
\end{align}
Unlike the dissipation term $\langle \mathcal{D}_{B}\rangle$, the production terms $\langle \mathcal{P}_{B1}\rangle$ and $\langle \mathcal{P}_{B2}\rangle$ are not sign-definite and so may in fact act to oppose the growth of $\langle\|\bm{B}\|^2\rangle$ \footnote{Despite the misnomer, we refer to them as production terms in keeping with the standard terminology used for the production terms in the Reynolds stress equation that are also not sign-definite \citep{pope00}.}. We must therefore consider the sign of these terms in order to understand the role they play in governing $\langle\|\bm{B}\|^2\rangle$. It will be shown that the sign of $\langle \mathcal{P}_{B2}\rangle$ is intimately connected to the emergence of ramp-cliff structures in the scalar field, and we therefore first consider in view of \eqref{Beq} how these structures form, and then show how this impacts the sign of $\langle \mathcal{P}_{B2}\rangle$ relative to that of $\langle \mathcal{P}_{B1}\rangle$.

When a scalar field is driven by a mean scalar gradient, ramp-cliff structures emerge which are associated with the fluctuating gradients developing a skewness whose sign corresponds to the direction of the imposed mean scalar gradient \citep{holzer94,sreenivasan18,buaria20a}. To understand how this asymmetry arises from the equation for $\bm{B}$, we may consider the case where the PDF of the initial condition $\bm{B}(0)$ is an isotropic and symmetric function, and uncorrelated from $\bm{A}$. Writing $\bm{B}$ in terms of Cartesian components, the equation for $B_z\equiv \bm{B\cdot}\bm{e}_z$ is

\begin{align}
D_t{B}_z&=-B_x A_{xz}-B_y A_{yz}-(B_z-\beta) A_{zz}+(Pr Re)^{-1}\nabla^2B_z,
\end{align}
where subscripts $x$ and $y$ denote components in the horizontal directions of the flow. For an isotropic flow, the PDFs of $A_{xz}$ and $A_{yz}$ are symmetric. Therefore, given the symmetric initial condition for $\bm{B}$, the symmetry breaking responsible for the PDF of $B_z$ becoming skewed cannot come from the terms $-B_x A_{xz}-B_y A_{yz}$ (or $(Pr Re)^{-1}\nabla^2B_z$), but must come from $-(B_z-\beta) A_{zz}$. As we will show momentarily, the strongest symmetry breaking associated with this term is generated in the range $|B_z|\in[0,\beta)$ and so we focus on this range. In the range $|B_z|\in[0,\beta)$ we can write $-(B_z-\beta) A_{zz}=|B_z-\beta| A_{zz}$, and so $A_{zz}<0$ events drive $B_z$ towards negative values, while $A_{zz}>0$ events drive $B_z$ towards positive values. Since in an isotropic flow, the PDF of $A_{zz}$ is negatively skewed, then the term $|B_z-\beta| A_{zz}$ will generate larger negative values of $B_z$ than positive ones, and hence negative skewness. If the flow field were Gaussian, however, this mechanism would be absent. Nevertheless, random Gaussian flows also generate skewed PDFs for $B_z$ \citep{holzer94} and, therefore, there must be another mechanism responsible for this. This second mechanism arises from the fact that starting from the isotropic initial condition for $B_z(0)$ and in a flow where the PDF of $A_{zz}$ is symmetric, then statistically, $-(B_z(0)-\beta) A_{zz}$ will be larger in regions where $B_z(0)<0$ than in regions where $B_z(0)>0$. This means that $-(B_z(0)-\beta) A_{zz}$ will generate larger negative values of $B_z$ than positive ones, and hence negative skewness. This mechanism fundamentally arises in \eqref{Beq} due to the ability of the fluctuating production $-\bm{A}^\top\bm{\cdot B}$ and mean gradient production $\beta\bm{A}^\top\bm{\cdot}\bm{e}_z$ terms to act together or against each other, and skewness of the PDF will be generated in the direction for which the two terms act together. The same argument applied to the case $\gamma>0$ shows that in this case $B_z$ will be positively skewed, the opposite of the $\gamma<0$ case.

In view of this, the emergence of ramp-cliff structures is determined by the interplay between $-\bm{A}^\top\bm{\cdot B}$ and $\beta\bm{A}^\top\bm{\cdot}\bm{e}_z$, which are associated with the production terms $\mathcal{P}_{B1}$ and $\mathcal{P}_{B2}$ in \eqref{Bsquared}. It may therefore be anticipated that ramp-cliff structures are also relevant to understanding the signs of the average terms $\langle \mathcal{P}_{B1}\rangle$ and $\langle \mathcal{P}_{B2}\rangle$. To consider this, we begin by examining the behavior of $\langle \mathcal{P}_{B1}\rangle$ and $\langle \mathcal{P}_{B2}\rangle$ in the ``short-time regime'' for the case where scalars are introduced to a fully-developed turbulent flow with initial condition $\bm{B}(0)=\bm{0}$ (a situation that will be of relevance to the DNS shown later). Using the Kolmogorov timescale $\tau_\eta$, for $t\ll\tau_\eta$ we have $\bm{A}(t)=\bm{A}(0)+O(t/\tau_\eta)$, and inserting this into \eqref{Beq} yields the solution $\bm{B}(t)\sim \beta t\bm{A}^\top(0)\bm{\cdot}\bm{e}_z+O([t/\tau_\eta]^2)$ when $\bm{B}(0)=\bm{0}$. From this we obtain
\begin{align}
\langle \mathcal{P}_{B2}\rangle\sim \beta^2 t\langle\|\bm{A}^\top(0)\bm{\cdot}\bm{e}_z\|^2\rangle+O([t/\tau_\eta]^2),
\end{align}
and hence at short times $\langle \mathcal{P}_{B2}\rangle>0$. Using the same approach we can also derive
\begin{align}
\langle \mathcal{P}_{B1}\rangle\sim \beta^2 t^2\langle\mathcal{P}_{A1}\rangle+O([t/\tau_\eta]^3).
\end{align}
The invariant $\mathcal{P}_{A1}\equiv-\bm{A}^\top\bm{:}(\bm{A\cdot A})$ is the velocity gradient self-amplification term and it is positive on average \citep{tsinober00} so that it acts as a source term in the equation for $\partial_t\langle\|\bm{A}\|^2\rangle$ (see equation \eqref{A2eq}). As a result $\langle \mathcal{P}_{B1}\rangle>0$ at short times, but its contribution is sub-leading compared to that from the mean gradient production term $\langle \mathcal{P}_{B2}\rangle$.

The question is whether the sign of these production terms remains the same once the stationary regime $\partial_t\langle\|\bm{B}\|^2\rangle= 0$ has been attained where the ramp-cliff structures are fully developed. The production terms may be re-expressed using $\bm{B}=\|\bm{B}\|\bm{e}_B$ and index notation as 
\begin{align}
\langle \mathcal{P}_{B1}\rangle &=-\langle\|\bm{B}\|^2 (\bm{e_B\cdot e}_j) A_{ji}(\bm{e}_i\bm{\cdot e_B})\rangle,\\
\langle \mathcal{P}_{B2}\rangle &=\beta\langle\|\bm{B}\| (\bm{e_B\cdot e}_j) A_{ji}(\bm{e}_i\bm{\cdot e}_z)\rangle.
\end{align}
Written in this form it is clear that these terms will only have the same sign if $\bm{e}_i\bm{\cdot e_B}$ and $\bm{e}_i\bm{\cdot e}_z$ tend to have opposite signs. This in turn depends on the alignments of $\bm{e_B}$ and $\bm{e}_z$ which is connected to the formation of the ramp-cliff structures in the flow.

Since $\langle\phi\rangle=0$ then $\langle B_z\rangle=0$, because $\langle B_z\rangle=\langle \nabla_z \phi\rangle=\nabla_z\langle \phi\rangle=0$. Ramp-cliff structures are associated with $B_z$ having larger negative than positive values (when $\gamma<0$). However, in order for $\langle B_z\rangle=0$ to be satisfied, it must be the case that events where $B_z>0$ are more probable than those with $B_z<0$. Since $B_z=\|\bm{B}\|\bm{e_B}\bm{\cdot e}_z$, a higher probability of $B_z>0$ events corresponds to a higher probability of $\bm{e_B}\bm{\cdot e}_z>0$ events than $\bm{e_B}\bm{\cdot e}_z<0$ events. Due to this, the most probable configuration is that the signs of $\bm{e}_i\bm{\cdot e}_z$ and $\bm{e}_i\bm{\cdot e_B}$ will be the same, and therefore once ramp-cliff structures emerge in the field, the production terms $\langle \mathcal{P}_{B1}\rangle$ and $\langle \mathcal{P}_{B2}\rangle$ will have opposite signs.

In order for the stationary regime $\partial_t\langle\|\bm{B}\|^2\rangle= 0$ to be sustained, it must be that case that $\langle \mathcal{P}_{B1}\rangle+\langle \mathcal{P}_{B2}\rangle>0$. As will be shown later, unless $Pr Re$ is very small then we expect  $|\langle \mathcal{P}_{B1}\rangle|>|\langle \mathcal{P}_{B2}\rangle|$. From this it follows that we must have $\langle \mathcal{P}_{B1}\rangle>0$, and therefore according to the argument above we will have $\langle \mathcal{P}_{B2}\rangle<0$ in the stationary regime due to the ramp-cliff structures.

\subsection{Effect of $Pr$ on the importance of the mean scalar gradient production}\label{Effect_Re_Pr}

We now want to understand how the contribution of $\langle \mathcal{P}_{B2}\rangle$ in \eqref{BBeq_stationary} relative to $\langle \mathcal{P}_{B1}\rangle$ depends on $Pr$. To understand this it is helpful to first think about the analogous, but better understood role of the mean shear on the fluctuating velocity gradient $\bm{A}$ and its dependence on $Re$ in homogeneous turbulence. For a time-independent mean velocity $\langle\bm{u}\rangle=\mathcal{S}z\bm{e}_x$, the equation for $\langle\|\bm{A}\|^2\rangle$ when $Fr=\infty$ is obtained using a forcing term $\bm{F}=-\mathcal{S}z(\bm{e}_x\bm{\cdot}\bm{A}+2\mathcal{S}\bm{e}_z)$ 
\begin{align}
\frac{1}{2}\partial_t\langle\|\bm{A}\|^2\rangle &=\langle\mathcal{P}_{A1}\rangle+\langle\mathcal{P}_{A2}\rangle-\langle\mathcal{D}_A\rangle,\label{A2eq}
\end{align}
where $\mathcal{P}_{A1}\equiv-\bm{A}^\top\bm{:}(\bm{A\cdot A})$ is the nonlinear self-amplification term, $\mathcal{P}_{A2}\equiv \bm{A}\bm{:}\bm{\nabla F}$ is the production associated with the mean shear, and $\mathcal{D}_A\equiv Re^{-1}\|\bm{\nabla A}\|^2$ is the dissipation rate of $\|\bm{A}\|^2$. Note that the pressure gradient term does not appear in \eqref{A2eq} because $\langle \bm{A:}\bm{\nabla\nabla}p\rangle=0$ for an incompressible, homogeneous flow. 

If $\|\bm{A}(0)\|\ll\mathcal{S}$, then initially, almost all of the production comes from the mean-shear term $\langle \mathcal{P}_{A2}\rangle$, and the viscous term is subleading so that $\langle\|\bm{A}\|^2\rangle$ grows. As $\langle\|\bm{A}\|^2\rangle$ continues to grow, both production terms contribute until eventually the dissipation term $\langle \mathcal{D}_{A}\rangle$ becomes large enough to arrest the growth (assuming a steady-state will be attained), at which point $\langle\|\bm{A}\|^2\rangle$ reaches a constant value. Whether the mean-shear term remains important in this steady-state limit depends on the value of $\langle\|\bm{A}\|^2\rangle$ in the steady state, and this will in turn depend upon $Re$. To see this, using a mean-field argument we can estimate that
\begin{align}
\Bigg\vert\frac{\langle\mathcal{P}_{A2}\rangle}{\langle\mathcal{P}_{A1}\rangle}\Bigg\vert\sim \frac{\mathcal{S}}{\sqrt{\langle\|\bm{A}\|^2\rangle}}\equiv \Lambda_A.\label{A2_production_ratio_estimate}
\end{align}
In view of this, if the viscous term arrests the growth of $\langle\|\bm{A}\|^2\rangle$ so that at steady state we have $\Lambda_A\geq O(1)$, then the mean-shear production term will play an important role in \eqref{A2eq} in the steady state. As $Re$ is increased, $\langle\|\bm{A}\|^2\rangle$ will increase (because the production terms have more time to amplify $\langle\|\bm{A}\|^2\rangle$ before the viscous term arrests the growth), and for sufficiently large $Re$ we will have $\Lambda_A\ll1$, such that the mean-shear production term will be irrelevant in \eqref{A2eq} in the steady-state.

We now want to similarly understand when the mean scalar gradient production term will become irrelevant in the equation for $\langle\|\bm{B}\|^2\rangle$. To consider this, we again use mean-field  estimates to obtain
\begin{align}
\Big\vert\frac{\langle \mathcal{P}_{B2}\rangle}{\langle \mathcal{P}_{B1}\rangle}\Big\vert \sim \frac{\beta}{\sqrt{\langle\|\bm{B}\|^2\rangle}}\equiv \Lambda_B.
\end{align}
Since $\langle\|\bm{B}\|^2\rangle$ will increase with increasing $Pr Re$ (because starting from an initial condition with $\|\bm{B}(0)\|\ll \beta$ the production terms have more time to amplify $\langle\|\bm{B}\|^2\rangle$ before dissipative effects arrest the growth), then for a given $\beta$, $\Lambda_B$ will decrease as $Pr Re$ increases, implying the role of mean scalar gradient production term in the equation for $\langle\|\bm{B}\|^2\rangle$ will become negligible for $Pr Re\to \infty$. We also note that the parameter $\Lambda_B$ is equal to the inverse of the square-root of the Cox number that is used in \citet{salehipour15}.

Since the equation for $\bm{B}$ is linear, then for a passive scalar where $\bm{A}$ is independent of $\beta$, the parameter $\Lambda_B$ is actually independent of $\beta$ and only depends on $Pr Re$.  To show this we write the equation for $\bm{B}$ in operator form as $\mathscr{L}\{\bm{B}\}=\beta\bm{A}^\top\bm{\cdot}\bm{e}_z$, where the linear operator is $\mathscr{L}\{\,\}\equiv D_t-(Pr Re)^{-1}\nabla^2+\bm{A}^\top\bm{\cdot}$. Since the inverse of a linear operator is also linear we have $\bm{B}=\mathscr{L}^{-1}\{\beta\bm{A}^\top\bm{\cdot}\bm{e}_z\}=\beta\mathscr{L}^{-1}\{\bm{A}^\top\bm{\cdot}\bm{e}_z\}$. From this it follows that
\begin{align}
\sigma_B=\beta\sqrt{\langle\| \mathscr{L}^{-1}\{\bm{A}^\top\bm{\cdot}\bm{e}_z\}\|^2\rangle},
\end{align}
and hence $\Lambda_B\equiv \beta/\sigma_B$ is independent of $\beta$ for a passive scalar (except for the trivial requirement that $\beta\neq 0$). 

Even though the average of the mean scalar gradient production term will become negligible in the equation for $\langle\|\bm{B}\|^2\rangle$ when $\Lambda_B\to 0$, this term still nevertheless plays a crucial implicit role which it must since without it the fluctuating scalar gradients would decay. To see this more clearly we should consider the behavior of the filtered gradients which provide information about the scalar gradients at different scales.

We define the filtering operation for an arbitrary field quantity $\bm{Y}$ to be
\begin{align}
\widetilde{\bm{Y}}(\bm{x},t)\equiv \int_{\mathbb{R}^3}\mathcal{G}_\ell(\|\bm{x}-\bm{x}'\|)\bm{Y}(\bm{x}',t)\,d\bm{x}',
\end{align}
where $\mathcal{G}_\ell$ is an isotropic filter kernel with filtering lengthscale $\ell$ (the particular choice of kernel, e.g. a Gaussian or box function, is not important here). Applying this filtering operator to equation \eqref{phi_nond} and taking the gradient of the resulting equation leads to
\begin{align}
\widetilde{D}_t\widetilde{\bm{B}}&=-\widetilde{\bm{A}}^\top\bm{\cdot}\widetilde{\bm{B}}+(Pr Re)^{-1}\nabla^2\widetilde{\bm{B}}+\beta\widetilde{\bm{A}}^\top \bm{\cdot e}_z-\bm{\nabla\nabla\cdot}\bm{\tau}_\phi,\label{Beq_filter}
\end{align}
where $\widetilde{D}_t\equiv\partial_t+(\widetilde{\bm{u}}\bm{\cdot\nabla})$, and $\bm{\tau}_\phi\equiv\widetilde{\bm{u}\phi}-\widetilde{\bm{u}}\widetilde{\phi}$ is the sub-grid stress vector.

From \eqref{Beq_filter}, the equation governing $\partial_t\langle\|\widetilde{\bm{B}}\|^2\rangle$ can be constructed, and for a statistically stationary, homogeneous flow it reduces to
\begin{align}
0=-\langle \widetilde{\bm{B}}\bm{\cdot}\widetilde{\bm{A}}^\top\bm{\cdot}\widetilde{\bm{B}}\rangle-(Pr Re)^{-1}\langle\|\bm{\nabla}\widetilde{\bm{B}}\|^2\rangle+\beta\langle \widetilde{\bm{B}}\bm{\cdot}\widetilde{\bm{A}}^\top \bm{\cdot e}_z\rangle-\langle \widetilde{\bm{B}}\bm{\cdot}\bm{\nabla\nabla\cdot}\bm{\tau}_\phi\rangle.
\end{align}
For $\ell\gg\eta_B$, where $\eta_B$ is the Batchelor length scale, the dissipation term $(Pr Re)^{-1}\langle\|\bm{\nabla}\widetilde{\bm{B}}\|^2\rangle$ can be ignored because almost all of the scalar dissipation takes place at scales $\ell=O(\eta_B)$. Therefore, for $\ell\gg\eta_B$ we have the balance
\begin{align}
-\langle \widetilde{\bm{B}}\bm{\cdot}\widetilde{\bm{A}}^\top\bm{\cdot}\widetilde{\bm{B}}\rangle+\beta\langle \widetilde{\bm{B}}\bm{\cdot}\widetilde{\bm{A}}^\top \bm{\cdot e}_z\rangle\sim \langle \widetilde{\bm{B}}\bm{\cdot}\bm{\nabla\nabla\cdot}\bm{\tau}_\phi\rangle.\label{inertial_balance_B}
\end{align}
The term $\langle \widetilde{\bm{B}}\bm{\cdot}\bm{\nabla\nabla\cdot}\bm{\tau}_\phi\rangle$ will be positive because this term describes how fluctuations are transferred on average to the sub-grid gradients from the filtered gradients, analogous to the kinetic and scalar variance cascades which are downscale in three dimensions.

Using mean-field estimates similar to those used before, \[|\langle \widetilde{\bm{B}}\bm{\cdot}\widetilde{\bm{A}}^\top\bm{\cdot}\widetilde{\bm{B}}\rangle|\sim \sqrt{\langle\|\widetilde{\bm{A}}\|^2\rangle}\langle\|\widetilde{\bm{B}}\|^2\rangle,\]and\[|\beta\langle \widetilde{\bm{B}}\bm{\cdot}\widetilde{\bm{A}}^\top \bm{\cdot e}_z\rangle|\sim \beta\sqrt{\langle\|\widetilde{\bm{A}}\|^2\rangle}\sqrt{\langle\|\widetilde{\bm{B}}\|^2\rangle}.\]Therefore, at scales where $\widetilde{\Lambda}_B\equiv\beta/\sqrt{\langle\|\widetilde{\bm{B}}\|^2\rangle}$ is $\ll 1$, the balance reduces to
\begin{align}
-\langle \widetilde{\bm{B}}\bm{\cdot}\widetilde{\bm{A}}^\top\bm{\cdot}\widetilde{\bm{B}}\rangle\sim \langle \widetilde{\bm{B}}\bm{\cdot}\bm{\nabla\nabla\cdot}\bm{\tau}_\phi\rangle,
\end{align}
while at scales where $\widetilde{\Lambda}_B\gg 1$ the balance reduces to
\begin{align}
\beta\langle \widetilde{\bm{B}}\bm{\cdot}\widetilde{\bm{A}}^\top \bm{\cdot e}_z\rangle\sim \langle \widetilde{\bm{B}}\bm{\cdot}\bm{\nabla\nabla\cdot}\bm{\tau}_\phi\rangle.
\end{align}
Since $ \langle \widetilde{\bm{B}}\bm{\cdot}\bm{\nabla\nabla\cdot}\bm{\tau}_\phi\rangle>0$, then we must have $\beta\langle \widetilde{\bm{B}}\bm{\cdot}\widetilde{\bm{A}}^\top \bm{\cdot e}_z\rangle>0$ at scales where $\widetilde{\Lambda}_B\gg 1$ in order for the balance to be satisfied. Therefore, although $\lim_{\ell/\eta_B\to 0}\beta\langle \widetilde{\bm{B}}\bm{\cdot}\widetilde{\bm{A}}^\top \bm{\cdot e}_z\rangle\to\beta\langle {\bm{B}}\bm{\cdot}{\bm{A}}^\top \bm{\cdot e}_z\rangle$ is predicted to be negative due to the ramp-cliff structures, at scales where $\widetilde{\Lambda}_B\gg 1$ is satisfied then $\beta\langle \widetilde{\bm{B}}\bm{\cdot}\widetilde{\bm{A}}^\top \bm{\cdot e}_z\rangle>0$. Hence the role of this mean gradient term in the equation governing $\langle\|\widetilde{\bm{B}}\|^2\rangle$ changes with scale, providing a source for $\langle\|\widetilde{\bm{B}}\|^2\rangle$ at scales where $\widetilde{\Lambda}_B\gg 1$, and providing a sink for $\langle\|\widetilde{\bm{B}}\|^2\rangle$ at scales where $\widetilde{\Lambda}_B\ll 1$. 

Note that regardless of $Re$ or $Pr$, there will always be a range of scales where $\widetilde{\Lambda}_B\gg 1$ is satisfied because statistical homogeneity of the flow enforces that $\lim_{\ell/L\to\infty}\widetilde{\bm{B}}\to0$, i.e. for sufficiently large scales, $\widetilde{\bm{B}}$ is equivalent to the spatial average of $\bm{B}$, which is zero. Due to this, $\lim_{\ell/L\to\infty}\widetilde{\Lambda}_B\to\infty$, regardless of $Re$ or $Pr$.

\subsection{Impact of mean gradient production term on the scalar dissipation rate}\label{PSanomaly}
According to the equation
\begin{align}
\label{eq:bsq}
\frac{1}{2}\partial_t\langle\|\bm{B}\|^2\rangle=\langle \mathcal{P}_{B1}\rangle+\langle \mathcal{P}_{B2}\rangle-\langle \mathcal{D}_{B}\rangle,
\end{align}
in a non-steady regime the quantity $\langle\|\bm{B}\|^2\rangle$ will continue to grow when the total production term is positive $\langle \mathcal{P}_{B1}\rangle+\langle \mathcal{P}_{B2}\rangle>0$ until the dissipation term $\langle \mathcal{D}_{B}\rangle$ grows to a large enough value to arrest the growth and generate the steady state $\partial_t\langle\|\bm{B}\|^2\rangle=0$. As $PrRe$ is increased it will take longer for this steady state to be attained and hence the production terms will have longer to act, causing the steady-state value of $\langle\|\bm{B}\|^2\rangle$ to increase as $PrRe$ is increased. If the increase of $\langle\|\bm{B}\|^2\rangle$ is proportional to $PrRe$ then a dissipation anomaly for the scalar field will be established where $\langle\chi\rangle$ is independent of $PrRe$.

The fact that the mean gradient production term $\langle \mathcal{P}_{B2}\rangle$ is negative means that $\langle \mathcal{P}_{B1}\rangle$ is not able to amplify $\langle\|\bm{B}\|^2\rangle$ to as large a value as it would have done if the term $\langle \mathcal{P}_{B2}\rangle$ were negligible. However, the mean-field estimate given earlier suggests that the resistance to the growth of $\langle\|\bm{B}\|^2\rangle$ coming from the term $\langle \mathcal{P}_{B2}\rangle$ reduces with decreasing $\Lambda_B$ and will become negligible in the regime $\Lambda_B\ll1$. Due to this, then momentarily ignoring the de-localization effect discussed earlier (see \S\ref{IBR}), the rate at which $\langle\|\bm{B}\|^2\rangle$ grows with increasing $Pr Re$ will itself depend upon $PrRe$ until the regime $\Lambda_B\ll1$ is reached, growing more rapidly with increasing $PrRe$ when $\Lambda_B\ll1$ than when $\Lambda_B\geq O(1)$. This in turn would suggest that anomalous behavior for $\langle\chi\rangle$ could only occur once the regime $\Lambda_B\ll1$ has been reached. However, the de-localization effect means that even for $\Lambda_B\ll1$, $\langle\chi\rangle$ will still in fact depend upon $Pr$ unless $Re_\lambda$ is sufficiently large. 

\section{Theory: gradient dynamics in stably stratified turbulence}\label{TSST}

Having considered the case of passive scalars we now turn to consider stably stratified turbulence. We will see that some of the properties that are already present for passive scalars play an important role in understanding stratified turbulence, and in particular, the role of ramp-cliff structures and the mean gradient production term.

\subsection{Buoyancy acts as both a source and a sink for velocity gradients in stratified turbulence}

The only difference between the gradient dynamics of passive scalar turbulence and stratified turbulence is the buoyancy term in the equation for $\bm{A}$. For a statistically homogeneous flow, the equation governing $\langle\|\bm{A}\|^2\rangle$ reduces to
\begin{align}
\frac{1}{2}\partial_t\langle\|\bm{A}\|^2\rangle &=\langle\mathcal{P}_{A1}\rangle-Fr^{-2}\langle \mathcal{P}_{B2}\rangle-\langle \mathcal{D}_{A}\rangle+\langle\mathcal{P}_{A2}\rangle ,
\label{eq:A_squared_homogeneous}
\end{align}
from which the pressure gradient term has disappeared because $\langle \bm{A:}\bm{\nabla\nabla}p\rangle=0$ for an incompressible, homogeneous flow. 

As discussed earlier, in a flow where $\bm{B}(0)=\bm{0}$, for $t\ll\tau_\eta$ we have $\langle \mathcal{P}_{B2}\rangle>0$, meaning that the buoyancy term in \eqref{eq:A_squared_homogeneous} acts as a sink for $t\ll\tau_\eta$.  However, once the ramp-cliff structures form and the stationary regime $\partial_t\langle\|\bm{B}\|^2\rangle= 0$ is attained, $\langle \mathcal{P}_{B2}\rangle<0$ and therefore buoyancy acts as a source term, contributing to the growth of $\langle\|\bm{A}\|^2\rangle$. Kinematically, for an incompressible, statistically homogeneous flow, $\langle\|\bm{A}\|^2\rangle=2\langle\|\bm{S}\|^2\rangle$ \citep{betchov56}, where $\bm{S}$ is the strain-rate. Therefore, since the buoyancy term acts to increase $\langle\|\bm{A}\|^2\rangle$, then it also acts to increase $\langle\|\bm{S}\|^2\rangle$ and hence the average TKE dissipation rate $\langle\epsilon\rangle=2Re^{-1}\langle\|\bm{S}\|^2\rangle$. Moreover, we also have the kinematic result $\langle\|\bm{A}\|^2\rangle=\langle\|\bm{\omega}\|^2\rangle$ \citep{betchov56}, where $\bm{\omega}$ is the vorticity, so that buoyancy also acts to increase enstrophy in the flow.

This conclusion seems surprising, because in stably stratified turbulence, buoyancy is expected to play the role of a sink term for turbulence. To understand the role of buoyancy on the velocity gradients in more detail we can use the filtering approach introduced earlier. Applying the filtering operator to equation \eqref{u_nond} and taking the gradient of the resulting equation yields
\begin{align}
\widetilde{D}_t\widetilde{\bm{A}}&=-\widetilde{\bm{A}}\bm{\cdot}\widetilde{\bm{A}}-\bm{\nabla\nabla}\widetilde{p}+Re^{-1}\nabla^2\widetilde{\bm{A}}-\beta Fr^{-2}\widetilde{\bm{B}}\bm{e}_z+\bm{\nabla}\widetilde{\bm{F}}-\bm{\nabla\nabla\cdot}\bm{\tau_u},\label{Aeq_filter}
\end{align}
where $\bm{\tau_u}\equiv\widetilde{\bm{uu}}-\widetilde{\bm{u}}\widetilde{\bm{u}}$ is the sub-grid stress tensor. From \eqref{Aeq_filter}, the equation governing $\partial_t\langle\|\widetilde{\bm{A}}\|^2\rangle$ can be constructed, and for a statistically stationary, homogeneous flow it reduces to
\begin{align}
0=-\langle\widetilde{\bm{A}}\bm{:}(\widetilde{\bm{A}}\bm{\cdot}\widetilde{\bm{A}})\rangle-
Re^{-1}\langle\|\bm{\nabla}\widetilde{\bm{A}}\|^2\rangle-\beta Fr^{-2}\langle \widetilde{\bm{B}}\bm{\cdot}\widetilde{\bm{A}}^\top \bm{\cdot e}_z\rangle+\langle\widetilde{\bm{A}}\bm{:}\bm{\nabla}\widetilde{\bm{F}}\rangle-\langle\widetilde{\bm{A}}\bm{:}\bm{\nabla\nabla\cdot}\bm{\tau_u}\rangle.
\end{align}
Once again, the pressure gradient term does not appear because $\langle \widetilde{\bm{A}}\bm{:\nabla\nabla}\widetilde{p}\rangle=0$ for an incompressible, homogeneous flow, assuming that the filtering operator is independent of position. The term $\langle \widetilde{\bm{A}}\bm{:}\bm{\nabla\nabla\cdot}\bm{\tau}_\phi\rangle$ will be positive because this term describes how fluctuations are transferred on average to the sub-grid gradients from the filtered gradients, analogous to the kinetic energy cascades which are downscale in three dimensions.

For $\ell\gg\eta$ the dissipation term $Re^{-1}\langle\|\bm{\nabla}\widetilde{\bm{A}}\|^2\rangle$ can be ignored because almost all of the dissipation takes place at scales $\ell=O(\eta)$, leading to the reduced balance
\begin{align}
\langle\widetilde{\bm{A}}\bm{:}\bm{\nabla}\widetilde{\bm{F}}\rangle\sim\langle\widetilde{\bm{A}}\bm{:}(\widetilde{\bm{A}}\bm{\cdot}\widetilde{\bm{A}})\rangle+\beta Fr^{-2}\langle \widetilde{\bm{B}}\bm{\cdot}\widetilde{\bm{A}}^\top \bm{\cdot e}_z\rangle+\langle\widetilde{\bm{A}}\bm{:}\bm{\nabla\nabla\cdot}\bm{\tau_u}\rangle.\label{inertial_balance}
\end{align}
Using the mean-field estimates\[|\langle\widetilde{\bm{A}}\bm{:}(\widetilde{\bm{A}}\bm{\cdot}\widetilde{\bm{A}})\rangle|\sim \langle\|\widetilde{\bm{A}}\|^2\rangle^{3/2},\]
and \[|\beta\langle \widetilde{\bm{B}}\bm{\cdot}\widetilde{\bm{A}}^\top \bm{\cdot e}_z\rangle|\sim \beta\sqrt{\langle\|\widetilde{\bm{A}}\|^2\rangle}\sqrt{\langle\|\widetilde{\bm{B}}\|^2\rangle},\]then in the regime $\sqrt{\langle\|\widetilde{\bm{A}}\|^2\rangle}\leq O\Big(\sqrt{\langle\|\widetilde{\bm{B}}\|^2\rangle}\Big)\ll\beta$ (which also implies $\widetilde{\Lambda}_B\gg 1$), the balance in \eqref{inertial_balance} reduces to
\begin{align}
\langle\widetilde{\bm{A}}\bm{:}\bm{\nabla}\widetilde{\bm{F}}\rangle\sim\beta Fr^{-2}\langle \widetilde{\bm{B}}\bm{\cdot}\widetilde{\bm{A}}^\top \bm{\cdot e}_z\rangle+\langle\widetilde{\bm{A}}\bm{:}\bm{\nabla\nabla\cdot}\bm{\tau_u}\rangle,\label{large_scale_balance}
\end{align}
and $\beta Fr^{-2}\langle \widetilde{\bm{B}}\bm{\cdot}\widetilde{\bm{A}}^\top \bm{\cdot e}_z\rangle>0$ for $\sqrt{\langle\|\widetilde{\bm{B}}\|^2\rangle}\ll\beta$, as shown earlier. This represents the balance at relatively large-scales where the production term due to forcing is balanced by losses due to buoyancy and transfer to smaller scales (which is analogous to the TKE equation \eqref{KE} because the TKE is dominated by the large-scales in high Reynolds number flows). The role of the buoyancy term $-\beta Fr^{-2}\langle \widetilde{\bm{B}}\bm{\cdot}\widetilde{\bm{A}}^\top \bm{\cdot e}_z\rangle$ is therefore subtle, opposing the production of velocity gradients at scales where $\widetilde{\Lambda}_B\gg 1$, but aiding their production at scales where $\widetilde{\Lambda}_B\ll 1$. This must be connected to the observation in \citet{legaspi20} based on their numerical simulations that the buoyancy spectrum changes sign and indicates transfer of potential energy to kinetic energy at high-wavenumbers in stratified turbulence. An investigation into this will be the subject of future work.

Note that in a flow where $\widetilde{\Lambda}_B\geq O(1)\,\forall \ell$ then the buoyancy term will act as a sink term for the velocity gradients at all scales, and corresponds to the case where buoyancy quenches turbulence at all scales.

\subsection{Prandtl number dependence of the kinetic and potential energy dissipation rates}\label{WCL}

Having considered how stratification impacts the velocity gradients through the buoyancy term, we now want to understand the impact of varying $Pr$ for a given $Re$. Analytical investigations into how $Pr$ impacts the velocity and density gradients are very difficult in general. However, we can obtain some insights based on a weak-coupling expansion to understand the impact of varying $Pr$ in the limit where the buoyancy term in the equation for $\bm{A}$ is weak (this does not assume that the role of buoyancy in the equation for $\bm{u}$ is weak). 

Weak-coupling expansions have been used for the Navier-Stokes equation in the context of renormalized perturbation theories. There the idea is to insert a non-dimensional coupling constant into the nonlinear term, and then expand the solutions in this constant. Although the constant is equal to 1 in the true problem, using it as an expansion parameter allows a regular perturbation expansion to be set up which can then be renormalized using methods developed in quantum field theory \citep{mccomb,McComb2002}. In that context, the coupling constant is an expansion parameter for the nonlinear term of Navier-Stokes, and the coupling it is associated with is scale coupling due to nonlinearity. In the present context where it is the role of buoyancy, not nonlinearity, that we want to understand, then the coupling constant should be inserted into the buoyancy term, and can be used as a parameter that controls the coupling between the velocity gradient and density gradient fields. To do this, we insert into the buoyancy term of equation \eqref{Aeq} a non-dimensional parameter $\lambda$, with $\lambda\to 0$ corresponding to the passive scalar limit. Conceptually, $\lambda$ could be thought of as an inverse buoyancy Reynolds number, reflecting the fact that in the limit of infinite buoyancy Reynolds number (i.e. $\lambda\to 0$), the effect of buoyancy on the velocity gradients should vanish. Note that we expand in $\lambda$ not for example $Fr$ because we are considering the case where the role of buoyancy in the equation for $\bm{A}$ is weak, even though its role in the equation for $\bm{u}$ may be strong (corresponding to small $Fr$).

Since equation \eqref{Aeq} with $\lambda$ inserted into the buoyancy term is regular in the limit $\lambda\to 0$, we introduce the perturbation expansions $\bm{A}=\sum_p \bm{A}_{(p)}\lambda^p$ and $\bm{B}=\sum_p \bm{B}_{(p)}\lambda^p$, where $\bm{A}_{(0)}$ and $\bm{B}_{(0)}$ are the solutions to equations \eqref{Aeq} and \eqref{Beq} for $\lambda=0$, i.e. the passive scalar case. Inserting the expansions into the expression for the total production term in the equation for $\langle\|\bm{A}\|^2\rangle$, then in the weak-coupling regime $\lambda\ll 1$ we obtain 
\begin{align}
\begin{split}
\langle \mathcal{P}_{A1}\rangle-Fr^{-2}\langle \mathcal{P}_{B2}\rangle &=-\langle\bm{A}_{(0)}^\top\bm{:}(\bm{A}_{(0)}\bm{\cdot}\bm{A}_{(0)}) \rangle -\beta Fr^{-2}\langle \bm{B}_{(0)}\bm{\cdot}(\bm{A}_{(0)}^\top\bm{\cdot}\bm{e}_z)\rangle +O(\lambda)\\
&\sim\sigma_{A0}^3+\beta Fr^{-2}\sigma_{A0}\sigma_{B0},
\end{split}
\end{align}
where $\sigma_{A0}\equiv \sqrt{\langle\|\bm{A}_{(0)}\|^2\rangle}$ and  $\sigma_{B0}\equiv \sqrt{\langle\|\bm{B}_{(0)}\|^2\rangle}$.

Since $\bm{A}_{(0)}$ corresponds to the solution for an unstratified flow then $\sigma_{A0}$ is independent of $Pr$, and $\sigma_{B0}$ will be an increasing function of $Pr$, with $\sigma_{B0}\propto \sqrt{Pr}$ if $Re$ and $Pr$ are in regimes where the passive scalar dissipation rate exhibits anomalous behavior. Consequently, in the weak-coupling regime the buoyancy term causes the total production term in the equation for $\langle\|\bm{A}\|^2\rangle$ to increase as $Pr$ increases for fixed $\beta$, and this suggests that $\langle\|\bm{A}\|^2\rangle$ will therefore also increase. For fixed $Re$ this in turn implies that $\langle\epsilon\rangle$ will increase as $Pr$ is increased.

Similarly, in the equation for $\langle\|\bm{B}\|^2\rangle$, the total production term in the weak-coupling regime is 
\begin{align}
\begin{split}
\langle \mathcal{P}_{B1}\rangle+\langle \mathcal{P}_{B2}\rangle&=-\langle\bm{B}_{(0)}\bm{\cdot}(\bm{A}_{(0)}^\top\bm{\cdot}\bm{B}_{(0)})\rangle+\beta\langle \bm{B}_{(0)}\bm{\cdot}(\bm{A}_{(0)}^\top\bm{\cdot}\bm{e}_z)\rangle+O(\lambda)\\
&\sim \sigma_{A0}\sigma_{B0}^2-\beta\sigma_{A0}\sigma_{B0}.
\end{split}
\end{align}
Since $\sigma_{A0}$ is independent of $Pr$, then the behavior of this total production term in the weak-coupling regime is the same that for passive scalar case that was discussed in \S\ref{PSanomaly}. In particular, the oppositional effect of $\langle \mathcal{P}_{B2}\rangle$ on the growth of $\langle\|\bm{B}\|^2\rangle$, together with the de-localization effect discussed in \S\ref{IBR}, implies that $\langle\chi\rangle$ will decrease with increasing $Pr$ unless $Re_\lambda$ is sufficiently large and $\beta/\sigma_{B0}$ is sufficiently small. 

Together with the earlier conclusion that $\langle\epsilon\rangle$ will increase with increasing $Pr$, the decrease of $\langle\chi\rangle$ with increasing $Pr$ means that in the weak-coupling regime the mixing coefficient $\Gamma\equiv Fr^{-2}\langle\chi\rangle/\langle\epsilon\rangle $ should decrease with increasing $Pr$. This prediction qualitatively agrees with the DNS results in \citet{riley23}.

Beyond the weak-coupling regime $\lambda\ll 1$, it is not possible to explore analytically the effect of $Pr$ on the velocity and density gradient dynamics without either renormalizing the expansion in $\lambda$ or else introducing closure approximations. How it behaves outside of the weak-coupling regime depends essentially on how the buoyancy term $-Fr^{-2}\langle \mathcal{P}_{B2}\rangle$ behaves. Provided that the parameter regime is such that $-Fr^{-2}\langle \mathcal{P}_{B2}\rangle$ remains positive, then the mean-field estimate $-\langle\mathcal{P}_{B2}\rangle\sim \beta \sigma_{A}\sigma_{B}$ suggests that even outside of the weak-coupling regime, $-Fr^{-2}\langle \mathcal{P}_{B2}\rangle$ will grow with increasing $Pr$ since the nature of the equation governing $\bm{B}$ essentially guarantees that the magnitude of the fluctuations of $\bm{B}$ (and therefore $\sigma_B$) increase with increasing $Pr$. In such a case, $\langle\epsilon\rangle$ will increase with increasing $Pr$ while $\langle\chi\rangle$ will decrease. Therefore, the predictions from the weak-coupling regime should carry over qualitatively to the case where the effects of buoyancy on $\bm{A}$ is not peturbative.

\subsection{The appropriate definition of the buoyancy Reynolds number}\label{Rb_def}

The analysis just presented suggests that the relative sizes of the buoyancy and inertial forces in the equation for $\langle\|\bm{A}\|^2\rangle$ will depend upon $Pr$. This has significant implications for whether the buoyancy Reynolds number can be used to reliably estimate the impact of buoyancy on the smallest scales of the flow, and the question of in which parameter regimes the behavior of the velocity and density gradients in stratified turbulence approach those of passive scalars.

\citet{riley03} proposed a buoyancy Reynolds number $Re_b\equiv Re Fr_N^2$ (recall $Fr_N=Fr/\beta$), which is expected to be proportional to the activity parameter \citep{gibson80} if both quantities are sufficiently large. In particular, \citet{riley03} developed a scaling analysis to estimate when the local gradient Richardson number will be less than one, and their analysis showed that this will be satisfied when $Re_b>1$.  When $Re_b>O(1)$ it is usually assumed that the effect of buoyancy on the smallest flow scales will be sub-leading, and negligible when $Re_b\gg 1$. In terms of the equation for $\bm{A}$, a more direct measure of the importance of buoyancy on the smallest flow scales is given by estimating the ratio of the square of the nonlinear to buoyancy terms in the equation for $\bm{A}$
\begin{align}
\mathcal{R}_b\equiv\Bigg\vert\frac{L^4 U^{-4}\langle\|\bm{A\cdot A}\|^2\rangle}{L\langle \|\beta Fr^{-2}\bm{B}\bm{e}_z\|^2\rangle}\Bigg\vert\sim \frac{Fr^4 L^3{\sigma_A^4}}{\beta^2 U^4\sigma_B^2}.\label{Rcalb}
\end{align}
Note that since $\mathcal{R}_b$ is a non-dimensional parameter, then to avoid confusion in the discussion that follows we have momentarily expressed $\bm{A}$ and $\bm{B}$, and all variables on which they depend, in dimensional form (which we do only in this sub-section). The other difference is that whereas in the rest of the paper $U,L$ are for simplicity based on values at some reference time, in this section they are defined instantaneously. As such, here $Re, Fr, Re_b$ and now $\mathcal{R}_b$ are all functions of time $t$ in general, which is in fact how they would often be defined in non-stationary flows, e.g. in decaying stratified turbulence \citep{riley23}.

For a homogeneous turbulent flow, when expressed in dimensional form we have $\sigma_A^2=\nu^{-1}\langle\epsilon\rangle$, $\sigma_B^2={(\nu/Pr)^{-1} \langle\chi\rangle}$,  $Fr^{-2} U^2L^{-1}\langle\chi\rangle$ is the dissipation rate of potential energy (the factor $U^2L^{-1}$ arises because the dimensions of $\bm{B}$ are the square root of an inverse length), and the mixing coefficient is $\Gamma\equiv Fr^{-2}\langle\chi\rangle/\langle\epsilon\rangle$. Using these relations in \eqref{Rcalb}, together with $Fr_N=Fr/\beta$, we obtain
\begin{align}
\mathcal{R}_b\sim \frac{Re_b }{Pr\Gamma}  \frac{\left<\epsilon\right> L}{ U^{3}}.
\label{trueRb}
\end{align}
In view of this, one significant difference between $\mathcal{R}_b$ and $Re_b$ is that the former explicitly depends on $Pr$ while the latter does not. Therefore, using $Re_b$ to estimate the importance of buoyancy on the small-scale gradient fields may not be reliable when $Pr$ deviates significantly from unity. 

In situations where Taylor's scaling $L\langle\epsilon\rangle/U^3\sim O(1)$ provides a reasonable estimate, $\mathcal{R}_b=O(Re_b)$ when $Pr=O(1)$ if $\Gamma=O(1)$, such that the standard buoyancy Reynolds number gives a reasonable estimate for the importance of buoyancy forces in the equation for $\bm{A}$. However, when $Pr> O(1)$ this need not be the case. Indeed, for increasing $Pr$, unless $\langle\epsilon\rangle L/(U^3\Gamma)$ increases faster than $\propto Pr$, then \eqref{trueRb} implies that $\mathcal{R}_b$ will become increasingly smaller than $Re_b$, such that the effects of buoyancy on the velocity gradients become increasingly strong as $Pr$ is increased.  

Whether this distinction between $Re_b$ and $\mathcal{R}_b$ matters in practice as a way of gauging the impact of buoyancy on the smallest flow scales depends upon the relevant ranges of $Re_b$ and $Pr$. For example, if $Re_b\ggg 1$, then unless $Pr$ is very large, we will also have $\mathcal{R}_b\ggg 1$. In this case having $Re_b\ggg 1$ would lead to the correct conclusion that the effects of buoyancy on the velocity gradients are negligible. For temperature stratified air and water $Pr\leq O(10)$, and over this range then provided $Re_b\gg 1$, $\mathcal{R}_b$ will likely also be large enough for the effects of buoyancy on $\bm{A}$ and $\bm{B}$ to be small. However, for salt-stratified water $Pr=O(1000)$, and this may cause $\mathcal{R}_b$ to be small enough for the effects of buoyancy on $\bm{A}$ to be important even when $Re_b\gg1$. Moreover, field observations in oceanic stratified flows show that $Re_b$ has a large range of values, spanning $O(10^{-2})\leq Re_b\leq O(10^5)$ (see figure 14 of \citet{jackson14}). This, together with the relevant ranges of $Pr$ indicates that in oceanic contexts, the difference between $Re_b$ and $\mathcal{R}_b$ may be significant, and therefore $\mathcal{R}_b$ should be used to determine the importance of buoyancy on the smallest flow scales rather than $Re_b$, since the latter does not correctly capture the impact of $Pr$ on the importance of buoyancy on the velocity gradient dynamics.

\section{Direct Numerical Simulations}
Data sets from direct numerical simulations (DNSs) will be used to explore the predictions and insights from the theoretical analysis. The first is a DNS of passive scalars which was previously reported in \citet{shete20} and \citet{shete22}. Specifically, we look at the DNS denoted in those papers as R633, with Taylor Reynolds number of 633, and $Pr=0.1, 1,7$ are resolved using $8190^3$, $8190^3$, and $14256^3$ grid points, respectively. The velocity field is homogeneous and isotropic, and is forced to be very nearly statistically stationary as described later in this section. There is a constant mean scalar gradient in the $z$-direction so that the scalar field is homogeneous in all directions, and the statistics are independent of direction in the horizontal.  

The second data set is of stably stratified turbulence which was previously reported in \citet{debk19} and \citet{riley23}.  The velocity field is forced to achieve homogeneous and isostropic turbulence, and then allowed to decay until it exhibits power-law decay with Taylor Reynolds number of 335, at which time the density field is initialised with zero fluctuations and allowed to decay subject to buoyancy. Simulations with $Pr=1$ and $Pr=7$ are considered which are resolved using grids of size $8192^2\times 4096$ and $12288^2\times 6144$, respectively, and in each case the domain is twice as large in the horizontal than the vertical directions.

For all the simulations, the domain is triply periodic so that a Fourier spectral method can be used to evolve the flows in time with minimal phase or truncation errors.  Derivatives and addition are done in Fourier space, multiplication is done in real space, a third-order Runge-Kutta schema is used to advance the solutions in time, and dealiasing is done with a combination of phase shifting, spectral truncation, and alternating between the advective and conservative forms of the nonlinear terms.  

The simulations require the specification of $\bm{F}$ in \eqref{eq:eq1} either to maintain the velocity field in a quasi-stationary state (for the passive scalar cases) or to initialise the velocity field (for the stably stratified cases). $\bm{F}$ is specified using a spring-damper model developed by \citet{overholt98} and generalised for the stratified case in \citet{rao11}.  The technique efficiently converges the velocity field to a prescribed spectrum at low wave numbers.  

\section{Results \& discussion}
\FloatBarrier
\subsection{Passive scalars}

\begin{figure}
\includegraphics{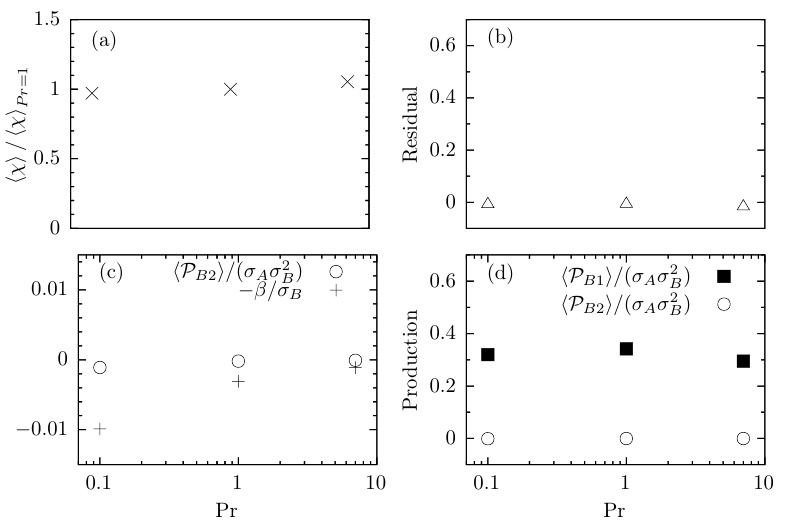}
\caption{Results for (a) $\langle\chi\rangle$ normalized by its value for $Pr=1$, (b) ``residual'' which is the sum of the r.h.s of \eqref{eq:bsq} normalized using $\sigma_A\sigma_B^2$, (c) $\langle\mathcal{P}_{B2}\rangle/(\sigma_A\sigma_B^2)$ compared with the mean-field prediction for this term, (d) production terms $\langle\mathcal{P}_{B1}\rangle$ and $\langle\mathcal{P}_{B2}\rangle$ both normalized using $\sigma_A\sigma_B^2$. Note that the same quantity plotted in (b) is termed ``unsteadiness'' for the decaying simulations (see figure \ref{fig:energy}).
\label{fig:passive}}
\end{figure}

We begin by considering results for passive scalars. In figure \ref{fig:passive}(a), the results for $\langle\chi\rangle$ as a function of $Pr$ are considered, normalized by the reference value at $Pr=1$, denoted by $\langle\chi\rangle_{Pr=1}$ (note that the vertical axis range used in the plot is chosen for fair comparison with the stratified results in figure \ref{fig:energy} for which this range is necessary). Over the range $Pr\in[0.1,7]$, $\langle\chi\rangle/\langle\chi\rangle_{Pr=1}$ increases with increasing $Pr$, with values going from approximately 0.97 to 1.05. To ensure that these variations are not due to a lack of stationarity of the scalar gradient field, in figure \ref{fig:passive}(b) we plot the ``residual'', which is the sum of the r.h.s of \eqref{eq:bsq}, scaled by the estimate for the production term, namely $\sigma_A\sigma_B^2$. The residual values are very small which indicates that the observed variations of $\langle\chi\rangle/\langle\chi\rangle_{Pr=1}$ are not due to a lack of small scale stationarity. 

The variations observed for $\langle\chi\rangle/\langle\chi\rangle_{Pr=1}$ in the passive scalar case would probably be considered negligible from a practical standpoint given that this variation corresponds to varying $Pr$ by two orders of magnitude. 
However, the variation could be considered non-negligible from a theoretical standpoint as it might indicate that $\langle\chi\rangle$ does not approach a constant as $Pr$ increases. In \S\ref{IBR} the model of \citet{donzis05} was discussed which in fact predicts that unless $Re_\lambda$ is sufficiently high, $L\langle\chi\rangle/(U\langle\phi^2\rangle)$ will vary with $Pr$ at a rate that is proportional to $1/\ln Pr$ for fixed $Re_\lambda$ and $Pr\geq 1$. DNS results in \citet{donzis05} confirmed this model prediction, as does the more recent study of \citet{Buaria21b} that considers the much larger value of $Re_\lambda=140$ with results spanning $Pr\in[1,512]$. Our results do not reveal such a strong $Pr$ dependence as theirs, but this is likely due to our DNS having the much higher value $Re_\lambda=633$, noting that the model of \citet{donzis05} predicts that $\langle\chi\rangle$ will become independent of $Pr$ (for finite $Pr$) in the limit $Re_\lambda\to \infty$. For our DNS with $Re_\lambda=633$, the model of \eqref{Donzis_model} predicts that the normalized dissipation rate $L\langle\chi\rangle/(U\langle\phi^2\rangle)$ will vary by $\approx 6\%$ in going from $Pr=1$ to $Pr=7$, and this is close to the magnitude of the variation that we observe. However, the model predicts that $L\langle\chi\rangle/(U\langle\phi^2\rangle)$ should decrease as $Pr$ increases; while our data shows that $L\langle\chi\rangle/(U\langle\phi^2\rangle)$ decreases in going from $Pr=0.1$ to $Pr=1$, it shows that it increases in going from $Pr=1$ to $Pr=7$. This discrepancy could be due to a lack of statistically stationarity of the large-scales of the passive scalar field in our DNS for $Pr=7$. Indeed, our DNS for $Re_\lambda=633$ and $Pr=7$ is extremely demanding computationally, and we are only able to construct the statistics by averaging over one large-eddy turnover time. This averaging window is much less than that used for the $Pr=0.1,1$ cases, and is also much less than that used in the DNS of \citet{Buaria21b} at the much lower value of $Re_\lambda=140$. Regardless of whether a lack of stationarity in the $Pr=7$ DNS explains the discrepancy or something else, what is far more important for the present study is that the variation of $\langle\chi\rangle/\langle\chi\rangle_{Pr=1}$ that we observe over the range $Pr\in[0.1,7]$ for passive scalars is very small compared to what is observed for stratified flows, as will be shown in \S\ref{SST}.

\begin{figure}
\includegraphics{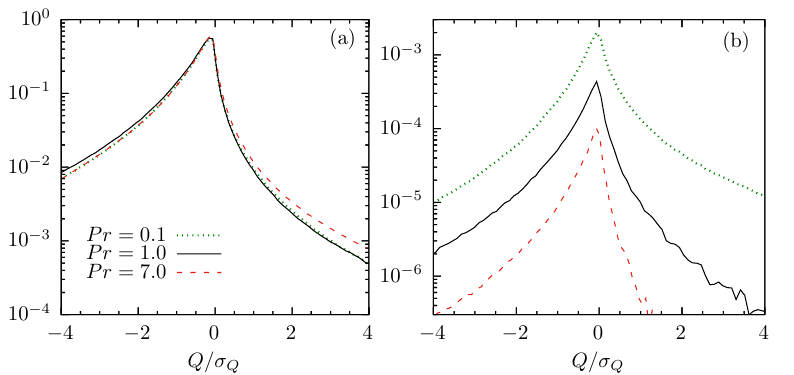}
\caption{Results for (a) $\sigma_A\sigma_B^{-2}\varphi(Q)\langle\mathcal{P}_{B1}\rangle_Q$ and (b) $-\sigma_A\sigma_B^{-2}\varphi(Q)\langle\mathcal{P}_{B2}\rangle_Q$ for the passive scalar cases. The horizontal axis is normalized using $\sigma_Q\equiv\sqrt{\langle Q^2\rangle}$.
\label{fig:passiveCavg}}
\end{figure}

In figure \ref{fig:passive}(b) we consider the mean production terms $\langle\mathcal{P}_{B1}\rangle$ and $\langle\mathcal{P}_{B2}\rangle$, scaled using the mean-field estimate for $\langle\mathcal{P}_{B1}\rangle$, namely $\sigma_A\sigma_B^2$. The value for  $\sigma_A^{-1}\sigma_B^{-2}\langle\mathcal{P}_{B1}\rangle$ at $Pr=1$ is very close to the value $0.32$ that has previously been reported from DNS at $Re_\lambda=250$ \citep{zhang_carbone_bragg_2023}. The scaled quantity $\sigma_A^{-1}\sigma_B^{-2}\langle\mathcal{P}_{B1}\rangle$ varies weakly with $Pr$, indicating that the mean-field estimate $\sigma_A\sigma_B^2$ accurately captures the dependence on $Pr$. The results for $\sigma_A^{-1}\sigma_B^{-2}\langle\mathcal{P}_{B2}\rangle$ show that this term is negligible compared with $\sigma_A^{-1}\sigma_B^{-2}\langle\mathcal{P}_{B1}\rangle$, and therefore it makes a negligible contribution to $\langle\chi\rangle$. Figure \ref{fig:passive}(c) compares $\sigma_A^{-1}\sigma_B^{-2}\langle\mathcal{P}_{B2}\rangle$ with the mean-field estimate for this term, namely $\sigma_A^{-1}\sigma_B^{-2}\langle\mathcal{P}_{B2}\rangle \sim -\beta/\sigma_B$. In agreement with the analysis in \S\ref{ERCS}, $\langle\mathcal{P}_{B2}\rangle$ is negative, which we argued is due to the emergence of ramp-cliff structures in the scalar field. Using mean-field estimates, it was argued in \S\ref{Effect_Re_Pr} that the magnitude of $\langle\mathcal{P}_{B2}\rangle$ should be negligible compared with $\langle\mathcal{P}_{B1}\rangle$ when $\Lambda_B\equiv \beta/\sigma_B\ll 1$. The results are consistent with this expectation, although figure \ref{fig:passive}(c) shows that the mean-field prediction $\sigma_A^{-1}\sigma_B^{-2}\langle\mathcal{P}_{B2}\rangle\sim -\beta/\sigma_B$ overestimates the magnitude of $\sigma_A^{-1}\sigma_B^{-2}\langle\mathcal{P}_{B2}\rangle$. It does, however, correctly predict that 
 the magnitude of $\sigma_A^{-1}\sigma_B^{-2}\langle\mathcal{P}_{B2}\rangle$ decays with increasing $Pr$.

A significant difference between the two production terms $\mathcal{P}_{B1}$ and $\mathcal{P}_{B2}$ relates to their behavior in rotation and strain dominated regions of the flow. In particular, using the strain-rate $\bm{S}$ and rotation-rate $\bm{R}$ decomposition $\bm{A}=\bm{S}+\bm{R}$ we have $\mathcal{P}_{B1}\equiv -\bm{B\cdot}\bm{A}^\top\bm{\cdot B}=-\bm{B\cdot}\bm{S}\bm{\cdot B}$ due to the antisymmetry of $\bm{R}$. Rotation therefore does not directly contribute to the fluctuating gradient production term $\mathcal{P}_{B1}$, but only indirectly contributes by influencing the alignments of $\bm{B}$ with respect to the eigenframe of $\bm{S}$. If we therefore conditionally average $\mathcal{P}_{B1}$ on the invariant $Q\equiv -\bm{A\cdot A}/2$, then we expect that the contribution to the average behavior
\begin{align}
\langle\mathcal{P}_{B1}\rangle=\int_\mathbb{R}\varphi(Q)\langle\mathcal{P}_{B1}\rangle_Q\,dQ,
\end{align}
(where $\varphi(Q)$ is the PDF of $Q$) from rotation (or vorticity) dominated regions $Q>0$ will be small compared with that from strain dominated regions $Q<0$. On the other hand, the rotation contribution to $ \langle \mathcal{P}_{B2}\rangle$, namely $\beta\langle \bm{B\cdot}(\bm{R}\bm{\cdot}\bm{e}_z)\rangle$, is not be zero because of the misalignment between $\bm{B}$ and $\bm{e}_z$. As a result, the contribution to the average behavior
\begin{align}
 \langle \mathcal{P}_{B2}\rangle=\int_\mathbb{R}\varphi(Q) \langle \mathcal{P}_{B2}\rangle_Q\,dQ,
\end{align}
from $Q>0$ regions may be significant compared with that from $Q<0$ regions. Taken together, this implies that the mean gradient production may play a much more significant role in governing $\|\bm{B}\|^2$ in rotation dominated regions than it does in strain dominated regions.

The results in figure \ref{fig:passiveCavg} for $\sigma_A\sigma_B^{-2}\varphi(Q)\langle\mathcal{P}_{B1}\rangle_Q$ show that this quantity is significantly skewed towards strain dominated regions where $Q<0$, and displays a weak dependence on $Pr$. This negative skewness comes entirely from $\langle\mathcal{P}_{B1}\rangle_Q$ because $\varphi(Q)$ is positively skewed in isotropic turbulence, which is associated with the vorticity field being more intermittent than the strain-rate field \citep{tsinober_book}. The implication is that the majority of the production associated with $\mathcal{P}_{B1}$ occurs in strain dominated rather than rotation dominated regions of the flow, as expected. For $\sigma_A\sigma_B^{-2}\varphi(Q)\langle\mathcal{P}_{B2}\rangle_Q$ the behavior is almost symmetric with respect to $Q$ for $Pr=0.1$, but becomes increasingly negatively skewed as $Pr$ increases. The values of $\sigma_A\sigma_B^{-2}\varphi(Q)\langle\mathcal{P}_{B2}\rangle_Q$ decrease dramatically as $Pr$ is increased (because of the reduction of $\beta/\sigma_B$ with increasing $Pr$), and at all $Pr$ considered the values are so small that there are no regions of the flow where $\mathcal{P}_{B2}$ plays a significant role in the production of the scalar gradients relative to $\mathcal{P}_{B1}$. From the mean-field estimates, this can again be understood as a consequence of the flows considered being in the regime where the parameter $\Lambda_B\equiv \beta/\sigma_B$ is very small. We will return later to consider $\sigma_A\sigma_B^{-2}\varphi(Q)\langle\mathcal{P}_{B2}\rangle_Q$ in the context of stratified flows, where its dependence on $Q$ can gives insights into how the buoyancy term $-Fr^{-2}\mathcal{P}_{B2}$ might behave differently in strain and rotation dominated regions of the flow.

\subsection{Stably stratified turbulence}\label{SST}

\begin{figure}
\includegraphics{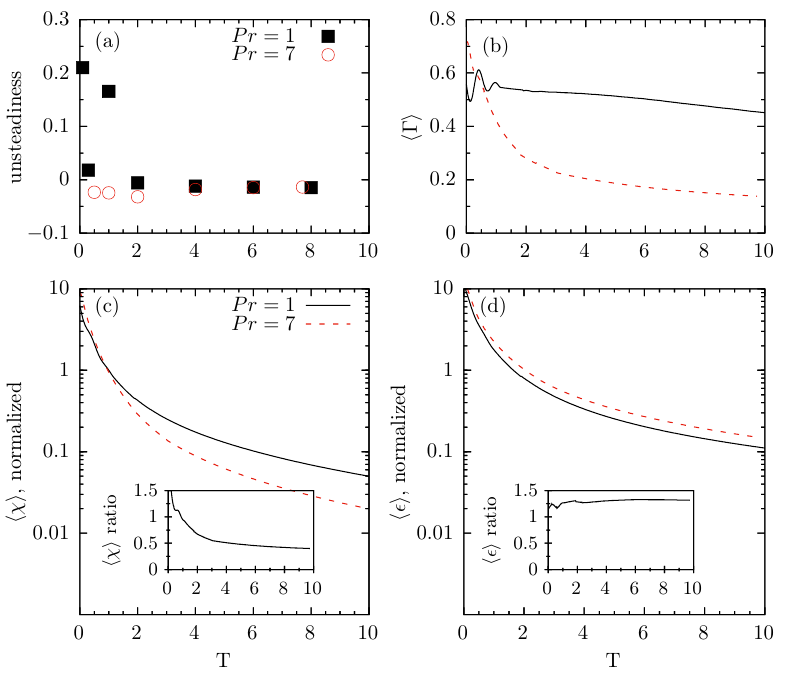}
\caption{Results for (a) ``unsteadiness'' which is $\partial_t\langle\|\bm{B}\|^2\rangle / (2 \sigma_A\sigma_B^2)$ computed via the sum of the terms on the rhs of \eqref{eq:bsq}, (b) mixing coefficient $\left<\Gamma\right> \equiv Fr^{-2} \left<\chi\right> / \left<\epsilon\right>$, (c) $\langle\chi\rangle$ normalized by its value for $Pr=1$ at $T=1$ (in order to be able to compare the effect of $Pr$ in the stratified case with that for the unstratified case shown in figure \ref{fig:passive}(a)), (d) $\langle\epsilon\rangle$ normalized by its value for $Pr=1$ at $T=1$. The inset plots in (c) and (d) show $\langle\chi\rangle_{Pr=7}/\langle\chi\rangle_{Pr=1}$ and $\langle\epsilon\rangle_{Pr=7}/\langle\epsilon\rangle_{Pr=1}$. In (b), the value for $\left<\Gamma\right>$ at $Pr=7$ and later times is consistent with that typically assumed for the ocean whereas the value for $Pr=1$ is much higher.  
\label{fig:energy}}
\end{figure}

We now turn to consider the results for stably stratified turbulence. One immediate difference between the DNS for passive scalars and stably-stratified turbulence is that in the former the large scales are quasi-stationary, whereas in the latter they are decaying. However, under Kolmogorov's quasi-equilibrium hypothesis we anticipate that the small-scales of the flow that dominate the velocity and scalar gradients will be in a state of quasi-equilibrium. To test this, in figure \ref{fig:energy}(a) we plot the sum of the terms on the rhs of the equation for $\partial_t\langle\|\bm{B}\|^2\rangle$ normalized by $\sigma_A\sigma_B^{2}$ (the mean-field estimate for $\langle \mathcal{P}_{B1}\rangle$), as a function of ``buoyancy time'' $T\equiv Nt/(2\pi)$. The results show that after an initial transient, $\sigma_A^{-1}\sigma_B^{-2}(1/2)\partial_t\langle\|\bm{B}\|^2\rangle$ becomes very small, indicating that the scalar gradients are indeed in a state of quasi-equilibrium. Therefore, the time dependence of the large-scale flow in the decaying stratified DNS should not cause significant differences for the scalar gradients compared with the passive scalar DNS, and any differences should be due to differences in the basic dynamics of the two cases.

In figure \ref{fig:energy}(b) we plot the mixing coefficient $\left<\Gamma\right> \equiv Fr^{-2} \left<\chi\right> / \left<\epsilon\right>$, and the results show that after the initial transient, $\langle\Gamma\rangle$ reduces dramatically as $Pr$ is increased from 1 to 7. Figure \ref{fig:energy}(c) and (d) show $\langle\chi\rangle$ and $\langle\epsilon\rangle$, respectively, normalized by their values for $Pr=1$ at $T=1$. The results show that as $Pr$ is increased from 1 to 7, $\langle\chi\rangle$ decreases while $\langle\epsilon\rangle$ increases. The insets in these plots show the ratios $\langle\chi\rangle_{Pr=7}/\langle\chi\rangle_{Pr=1}$ and $\langle\epsilon\rangle_{Pr=7}/\langle\epsilon\rangle_{Pr=1}$ in order to show more clearly the size of the variations. The results show that after the initial transient, $\langle\chi\rangle$ decreases by roughly $50\%$ as $Pr$ is increased from 1 to 7, while $\langle\epsilon\rangle$ increases by roughly $25\%$. This very strong reduction in $\langle\chi\rangle$ for stratified turbulence as $Pr$ is increased is in stark contrast to what was observed earlier for the passive scalar runs where $\langle\chi\rangle$ varied by only $\approx 6\%$ as $Pr$ is increased from 1 to 7. 

At $T=1.5$, when $\langle\chi\rangle$ has already dropped by $\approx 25\%$ in going from $Pr=1$ to $Pr=7$, the activity parameter $Gn\equiv \langle\epsilon\rangle/(\nu N^2)$ is $\approx 20$. This would usually be taken to suggest that buoyancy is playing a sub-leading role in the behavior of the small-scale gradients that govern $\langle\chi\rangle$, and that the scalar gradients behave like those for a passive scalar. If this is the case, then the model of \citet{donzis05} should apply, according to which $\langle\chi\rangle$ will decrease with increasing $Pr$ for $Pr\geq 1$ due to the emergence of the viscous-convective range, unless $Re_\lambda$ is sufficiently high. Since the value of $Re_\lambda$ in our DNS of stratified turbulence is much smaller (at $T=0$, $Re_\lambda=335$) than that in the DNS of passive scalars shown earlier (where $Re_\lambda=633$), perhaps the much stronger $Pr$ dependence of $\langle\chi\rangle$ for the stratified runs compared with the passive scalar runs is simply due to $Re_\lambda$ being much smaller in the former and not due to the effect of buoyancy. To test this we used the model of \cite{donzis05} with the values of $Re_\lambda$ in our DNS of stratified turbulence and found that their model predicts $\lesssim 14\%$ reduction of $\langle\chi\rangle$ (the reduction predicted depends on time since $Re_\lambda$ is a function of time in the stratified flow) in going from $Pr=1$ to $Pr=7$. This variation is far smaller than the $\approx 50\%$ reduction we observe in figure \ref{fig:energy}(c). Hence, although the effect of the viscous-convection regime, which is captured in the model of \citet{donzis05}, may play a role in explaining why $\langle\chi\rangle$ reduces in our stratified flow when going from $Pr=1$ to $Pr=7$, it is certainly not the main cause.

\begin{figure}
\includegraphics{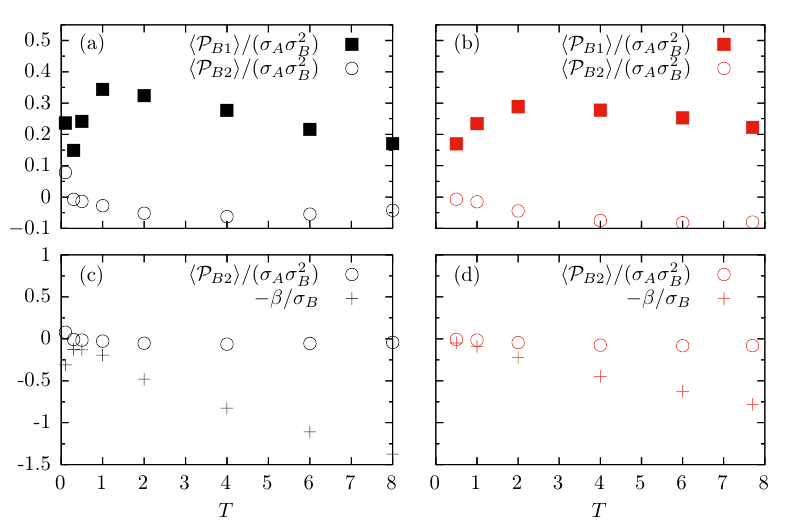}
\caption{Results for $\langle \mathcal{P}_{B1}\rangle$ and $\langle \mathcal{P}_{B2}\rangle$ normalized by $\sigma_A\sigma_B^2$ as a function of buoyancy time $T=Nt / (2\pi)$ for the stratified cases at (a) $Pr=1$ and (b) $Pr=7$.  The mean-field prediction for $\langle \mathcal{P}_{B2}\rangle/(\sigma_A\sigma_B^2)$ is tested via the lower two panels at (c) $Pr=1$ and (d) $Pr=7$. 
\label{fig:meanP_B}}
\end{figure}

According to the analysis of \S\ref{WCL}, a strong dependence of $\langle\chi\rangle$ on $Pr$ will arise when the mean gradient production term $\langle\mathcal{P}_{B2}\rangle$ plays a sufficiently large role in the equation governing $\langle\|\bm{B}\|^2\rangle$, and the fact that the dependence of $\langle\chi\rangle$ on $Pr$ is much stronger for the stratified case than for the passive scalar case must be due to $\langle\mathcal{P}_{B2}\rangle$ playing a much more significant role in the former case than the latter. To test this, in figure \ref{fig:meanP_B} we plot $\langle\mathcal{P}_{B1}\rangle$ and $\langle\mathcal{P}_{B2}\rangle$, normalized by $\sigma_A\sigma_B^2$. As for the passive scalar case, the results show that $\langle\mathcal{P}_{B1}\rangle$ is positive, meaning that the fluctuating gradient production term acts as a source for $\langle\|\bm{B}\|^2\rangle$. In agreement with the analysis of \S\ref{ERCS} (which also applies to the stratified case), the mean gradient production term $\langle \mathcal{P}_{B2}\rangle$ is positive at $t\ll\tau_\eta$ (this is only observable for the $Pr=1$ case; we do not have data at small enough $T$ for the $Pr=7$ case to observe it), but then becomes negative once the ramp-cliff structures have emerged. Following the arguments of \S\ref{WCL}, due to the presence of this term opposing the growth of $\langle\|\bm{B}\|^2\rangle$, when $Pr$ is increased from 1 to 7, $\langle \mathcal{P}_{B1}\rangle$ is not able to amplify $\langle\|\bm{B}\|^2\rangle$ to as large a value as it would have done were the term $\langle \mathcal{P}_{B2}\rangle$ negligible in the equation for $\langle\|\bm{B}\|^2\rangle$. While $\langle \mathcal{P}_{B2}\rangle$ is smaller in magnitude than $\langle \mathcal{P}_{B1}\rangle$ according to figure \ref{fig:meanP_B} (as it must be in order for the total production term to be positive), it is nevertheless significant and opposes the production of $\langle\|\bm{B}\|^2\rangle$. Furthermore, the results show that $-\sigma_A^{-1}\sigma_B^{-2}\langle\mathcal{P}_{B2}\rangle$ reaches values up to $\approx 0.1$, which is around two orders of magnitude larger than what is observed in figure \ref{fig:passive}(c) for the passive scalar case. By contrast, the values for $\sigma_A^{-1}\sigma_B^{-2}\langle\mathcal{P}_{B1}\rangle$ for the stratified case (after the initial transient) and passive scalar are comparable. This strongly supports the argument of \S\ref{WCL} that it is the contribution from $\langle\mathcal{P}_{B2}\rangle$ that is responsible for the strong reduction of $\langle\chi\rangle$ in stratified flows as $Pr$ is increased.

\begin{figure}
\includegraphics{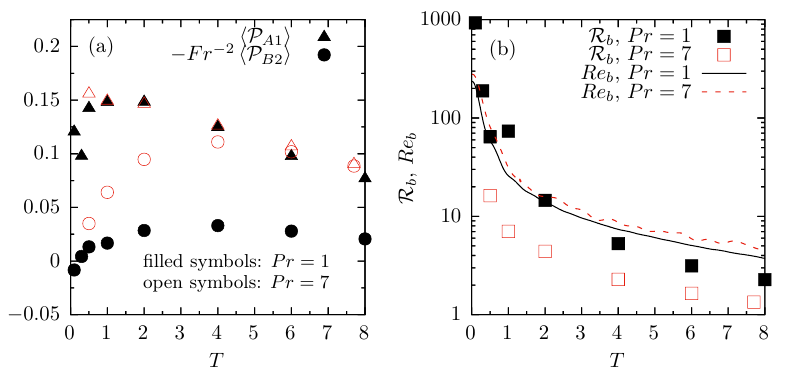}
\caption{Results from stratified DNS for (a) $\langle \mathcal{P}_{A1}\rangle$ and $-Fr^{-2}\langle \mathcal{P}_{B2}\rangle$ normalized by $\sigma_A^{3}$, (b) $Re_b$ and $\mathcal{R}_b$, the latter being computed based on the estimate on the r.h.s of equation \eqref{Rcalb}.
\label{fig:meanP_A}}
\end{figure}

While the impact of buoyancy on the momentum field in stratified turbulence doubtless plays an important role in causing the magnitude of $\sigma_A^{-1}\sigma_B^{-2}\langle\mathcal{P}_{B2}\rangle$ to be much larger for the stratified case than the passive scalar case, the analysis of \S\ref{WCL} suggests that the other key factor is the size of the non-dimensional parameter $\Lambda_B\equiv\beta/\sigma_B$, because the mean-field estimate suggests $\sigma_A^{-1}\sigma_B^{-2}\langle\mathcal{P}_{B2}\rangle\sim -\Lambda_B$. Comparing figure \ref{fig:meanP_B}(c) and (d) with figure \ref{fig:passive}(c) shows that $\Lambda_B\equiv\beta/\sigma_B$ is in fact two orders of magnitude larger in the stratified cases than the passive scalar cases. As for the passive scalar case, the mean-field prediction overestimates the magnitude of $\sigma_A^{-1}\sigma_B^{-2}\langle\mathcal{P}_{B2}\rangle$ in the stratified flows. However, whereas the mean-field prediction that $\sigma_A^{-1}\sigma_B^{-2}\langle\mathcal{P}_{B2}\rangle$ should decrease in magnitude as $Pr$ is increased was confirmed for passive scalars, it is not for the stratified case. Indeed, figures \ref{fig:meanP_B}(a) and (b) show that for $T\gtrsim 4$, $\sigma_A^{-1}\sigma_B^{-2}\langle\mathcal{P}_{B2}\rangle$ actually increases in magnitude as $Pr$ increases from 1 to 7. Therefore, the difference in the behavior of $\langle\mathcal{P}_{B2}\rangle$ in the passive scalar and stratified cases cannot be simply accounted for by the differences of the values of $\Lambda_B\equiv\beta/\sigma_B$ in these flows. The effects of buoyancy in the stratified flows, which are absent for the passive scalar, are clearly playing a key role in influencing $\langle\mathcal{P}_{B2}\rangle$ and its dependence on $Pr$. A significant implication of the finding that $\sigma_A^{-1}\sigma_B^{-2}\langle\mathcal{P}_{B2}\rangle$ increases with increasing $Pr$ is that it is not clear when $\langle\chi\rangle$ will become independent of $Pr$ in stratified turbulence. However, we do know that the effects of buoyancy at the small-scales must vanish in the limit $\mathcal{R}_b\to \infty$ (see \S\ref{Rb_def}), and according to the model of \citet{donzis05}, $\langle\chi\rangle$ will be independent of $Pr$ in the limit $Re_\lambda\to\infty$ for a passive scalar field.
In order to understand the increase of $\langle\epsilon\rangle$ with increasing $Pr$ observed in figure \ref{fig:energy}(d), in figure \ref{fig:meanP_A}(a) we plot $\langle \mathcal{P}_{A1}\rangle$ and $-Fr^{-2}\langle \mathcal{P}_{B2}\rangle$ normalized by $\sigma_A^{3}$ (the mean-field estimate for $\langle \mathcal{P}_{A1}\rangle$) for the stratified cases with $Pr=1$ and $Pr=7$. For $T\to 0$, $\sigma_A^{-3}\langle\mathcal{P}_{A1}\rangle$ is close to the value $0.15$ which has been observed for statistically stationary, isotropic turbulence \citep{bragg2022}. The results also show that $\sigma_A^{-3}\langle\mathcal{P}_{A1}\rangle$ only depends weakly on $Pr$, and therefore this term is not responsible for the $Pr$ dependence of $\langle\epsilon\rangle$ observed in figure \ref{fig:energy}(d), just as for the weak-coupling regime analyzed in \S\ref{WCL}. For $Pr=1$, the buoyancy term at short times is negative (the same should also occur for the $Pr=7$ case, but we do not have data at small enough $T$ to check this), $-Fr^{-2}\langle \mathcal{P}_{B2}\rangle<0$, in agreement with the asymptotic analysis of \S\ref{ERCS} that predicts $\langle \mathcal{P}_{B2}\rangle\sim \beta^{2} t\langle\|\bm{A}^\top(0)\bm{\cdot}\bm{e}_z\|^2\rangle\geq 0$ at $t/\tau_\eta\ll1$. This means that buoyancy makes a negative contribution to $\partial_t\langle\|\bm{A}\|^2\rangle$ at short times which is likely the reason why for the $Pr=1$ case $\sigma_A^{-3}\langle\mathcal{P}_{A1}\rangle$ initially reduces. However, in agreement with the arguments in \S\ref{ERCS}, the buoyancy term subsequently becomes positive, $-Fr^{-2}\langle \mathcal{P}_{B2}\rangle>0$, due to the emergence of the ramp-cliff structures in the flow and the associated alignments between $\bm{B}$ and the mean scalar gradient direction $\bm{e}_z$. Therefore, for both $Pr=1$ and $Pr=7$, buoyancy acts as a source term for $\langle\|\bm{A}\|^2\rangle$ after the initial transient.

The results in figure \ref{fig:meanP_A}(a) also show, in qualitative agreement with the predictions from the weak-coupling analysis in \S\ref{WCL}, that the scaled buoyancy term $-\sigma_A^{-3} Fr^{-2}\langle \mathcal{P}_{B2}\rangle$ increases with increasing $Pr$, causing $\langle\|\bm{A}\|^2\rangle$ to increase with increasing $Pr$. This then is the reason why in figure \ref{fig:energy}(d) $\langle\epsilon\rangle$ was observed to increase with increasing $Pr$, since for a homogeneous flow $\langle\epsilon\rangle=Re^{-1}\langle\|\bm{A}\|^2\rangle$. Moreover, whereas for $Pr=1$, $-\sigma_A^{-3} Fr^{-2}\langle \mathcal{P}_{B2}\rangle$ is considerably smaller than $\sigma_A^{-3}\langle\mathcal{P}_{A1}\rangle$, for $Pr=7$ they are almost equal. Therefore, remarkably, for $Pr=7$, buoyancy is playing as large a role in generating the velocity gradients as the combined processes of vortex stretching and strain self-amplification that are described by $\langle \mathcal{P}_{A1}\rangle$. Whether $-\sigma_A^{-3} Fr^{-2}\langle \mathcal{P}_{B2}\rangle$ will continue to grow as $Pr$ is further increased or whether it will saturate it not certain. The mean-field estimate is $-\langle \mathcal{P}_{B2}\rangle \sim \beta \sigma_{A}\sigma_{B}$, and as discussed earlier, the nature of the equation for $\bm{B}$ virtually guarantees that $\sigma_B$ will be an increasing function of $Pr$ (for a given $Re$). From this it follows that provided the buoyancy term continues to play the role of a source term in the equation for $\bm{A}$ as $Pr$ is increased (and there is no reason to think it will not), then $-Fr^{-2}\langle \mathcal{P}_{B2}\rangle$ will be an increasing function of $Pr$, which has profound implications for understanding mixing in stably stratified flows in regimes where $Pr$ is large (e.g. salt-stratified water flows). 

In the discussion so far, and also in the analysis in \S\ref{TSST}, we have used the kinematic result for incompressible, homogeneous turbulence $\langle\|\bm{A}\|^2\rangle=2\langle\|\bm{S}\|^2\rangle$ \citep{betchov56} to show that since buoyancy acts as a source term in the equation for $\langle\|\bm{A}\|^2\rangle$ and since this buoyancy contribution becomes stronger with increasing $Pr$, then this explains why $\langle\epsilon\rangle$ increases with increasing $Pr$. However, to demonstrate this on strictly dynamical grounds we ought to provide an explanation in terms of the effect of buoyancy on $\langle\|\bm{S}\|^2\rangle$, not on $\langle\|\bm{A}\|^2\rangle$, since from a fundamental perspective $\langle\epsilon\rangle\equiv 2Re^{-1}\langle\|\bm{S}\|^2\rangle$ for an incompressible Newtonian fluid such that rotational motion in the fluid (which is contained in $\langle\|\bm{A}\|^2\rangle$) plays no explicit role in the dissipation rate of TKE. The equation governing $\langle\|\bm{S}\|^2\rangle$ for a homogeneous turbulent flow with buoyancy is similar to that for $\langle\|\bm{A}\|^2\rangle$

\begin{align}
\begin{split}    
\frac{1}{2}\partial_t\langle\|\bm{S}\|^2\rangle =&-\langle\bm{S}\bm{:}(\bm{S\cdot S})\rangle-(1/4)\langle\bm{S}\bm{:}\bm{\omega\omega}\rangle-\beta Fr^{-2}\langle\bm{B\cdot}\bm{S}\bm{\cdot}\bm{e}_z \rangle \\
&-Re^{-1}\langle\|\bm{\nabla S}\|^2\rangle+\langle\bm{S}\bm{:}\bm{\nabla F}\rangle.
\end{split}
\label{S2eq}
\end{align}
It is straightforward to show that for an incompressible, homogeneous flow $\langle\bm{B\cdot}\bm{S}\bm{\cdot}\bm{e}_z \rangle =(1/2)\langle\bm{B\cdot}\bm{A}^\top\bm{\cdot}\bm{e}_z \rangle $, and therefore $-\beta Fr^{-2}\langle\bm{B\cdot}\bm{S}\bm{\cdot}\bm{e}_z \rangle=-(1/2)Fr^{-2}\langle\mathcal{P}_{B2}\rangle$. Consequently, just as we have demonstrated that the buoyancy term $-Fr^{-2}\langle\mathcal{P}_{B2}\rangle$ acts as a source term that causes $\langle\|\bm{A}\|^2\rangle$ to increase with increasing $Pr$, it is also the case that the buoyancy term $-\beta Fr^{-2}\langle\bm{B\cdot}\bm{S}\bm{\cdot}\bm{e}_z \rangle$ acts as a source term that causes $\langle\|\bm{S}\|^2\rangle$ to increase with increasing $Pr$. Hence, it is indeed the case that buoyancy is the dynamical cause of $\langle\epsilon\rangle$ increasing with increasing $Pr$. It is also worth mentioning that the relative size of the buoyancy term to the nonlinear term in the equation for $\langle\|\bm{S}\|^2\rangle$ is the same as that in the equation for $\langle\|\bm{A}\|^2\rangle$. This follows both because  $-\beta Fr^{-2}\langle\bm{B\cdot}\bm{S}\bm{\cdot}\bm{e}_z \rangle=-(1/2) Fr^{-2}\langle\mathcal{P}_{B2}\rangle$ and also because using the results from \citet{betchov56} it can be shown that $-\langle\bm{S}\bm{:}(\bm{S\cdot S})\rangle-(1/4)\langle\bm{S}\bm{:}\bm{\omega\omega}\rangle=-(1/2)\langle\mathcal{P}_{A1}\rangle$ for an incompressible, homogeneous flow.

In \S\ref{Rb_def} it was argued that the relative size of the inertial to buoyancy forces in the equation for $\bm{A}$ could be estimated using $\mathcal{R}_b$, and the results for this quantity are shown in figure \ref{fig:meanP_A}(b). Consistent with the results in figure \ref{fig:meanP_A}(a), the results show that $\mathcal{R}_b$ decreases as $Pr$ is increased, indicating that buoyancy plays an increasingly important role in the dynamics governing $\bm{A}$ as $Pr$ is increased. By contrast, the results in figure \ref{fig:meanP_A}(b) also show that the buoyancy Reynolds number $Re_b\equiv Re Fr_N^2$ (based on the instantaneous values of $Re$ and $Fr_N$) increases slightly in going from $Pr=1$ to $Pr=7$, which would incorrectly suggest that the impact of buoyancy on the dynamics of the smallest flow scales reduces as $Pr$ is increased. At $Pr=1$, the quantities $\mathcal{R}_b$ and $Re_b$ are, however, quite close. Moreover, the activity parameter $Gn\equiv \langle\epsilon\rangle/(\nu N^2)$, which is sometimes used as an alternative to $Re_b$, suffers from the same issue as $Re_b$, namely that it does not correctly capture the effect of $Pr$ on how buoyancy impacts the smallest scales of the flow. As seen earlier, $\langle\epsilon\rangle$ increases with increasing $Pr$, which would then imply that $Gn$ also increases. If $Gn$ is used as a metric to gauge the impact of buoyancy on the smallest scales of the flow, then this would imply that the impact of buoyancy on the smallest scales is smaller for $Pr=7$ than $Pr=1$, which is incorrect. Therefore, $\mathcal{R}_b$, rather than either $Re_b$ or $Gn$, should be used as the metric for estimating the importance of buoyancy on the smallest scales in a stratified flow.  

\begin{figure}
\includegraphics{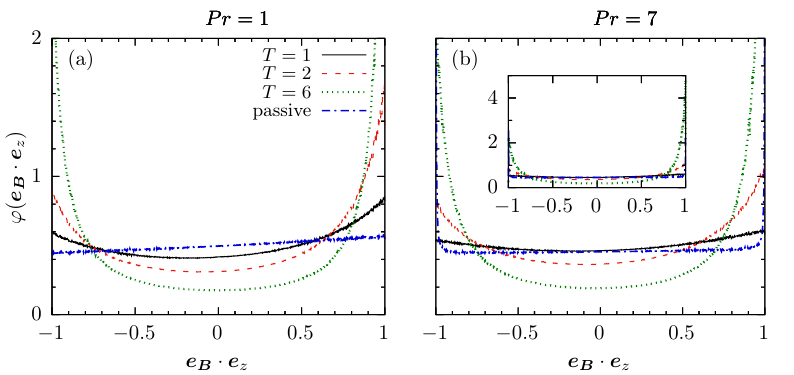}
\caption{Results for the probability density function of $\bm{e_B}\bm{\cdot e}_z$ for (a) $Pr=1$ and (b) $Pr=7$. Stratified results are shown for different buoyancy times $T$.} 
\label{fig:alignment}
\end{figure}

According to the argument presented in \S\ref{ERCS}, the reason why $\langle \mathcal{P}_{B2}\rangle$ transitions from being positive at $t\ll\tau_\eta$ to negative is due to the emergence of ramp-cliff structures in the flow which are associated with a preference for $\bm{e_B}\equiv\bm{B}/\|\bm{B}\|$ to be aligned with $\bm{e}_z$. More specifically, the argument is that $\langle \mathcal{P}_{B2}\rangle$ becoming negative is associated with $\bm{e_B}\bm{\cdot e}_z>0$ events being more probable than $\bm{e_B}\bm{\cdot e}_z<0$ events (when $\gamma<0$) due to the mechanism that generates the ramp-cliff structures. To test this, in figure \ref{fig:alignment} we plot the PDF of $\bm{e_B}\bm{\cdot e}_z$, namely $\varphi(\bm{e_B}\bm{\cdot e}_z)$, for the stratified flows as well as the passive scalar results for reference. The stratified results for $Pr=1$ show a clear bias towards $\bm{e_B}\bm{\cdot e}_z>0$ events, consistent with the argument in \S\ref{ERCS}. As $T$ increases the PDF reduces and becomes more uniform over the central region of the space $\bm{e_B}\bm{\cdot e}_z\in[-1,+1]$, while it increases and becomes less uniform closer to the edges of the space. This suggests that as $T$ increases and the flow becomes increasingly stratified, the conditions required for the generation of the ramp-cliff structures are only satisfied in extreme regions of the flow where the behavior of $\bm{B}$ differs strongly from its mean-field behavior. The results for $Pr=7$ show similar behavior except that the asymmetry of $\varphi(\bm{e_B}\bm{\cdot e}_z)$ is weaker than for $Pr=1$. Interestingly, however, the results in \citet{riley23} for the same data set show that the skewness of $B_z$ becomes stronger in going from $Pr=1$ and $Pr=7$. This difference reflects the fact that while the skewness of $B_z$ is directly connected to asymmetry in $\varphi(\bm{e_B}\bm{\cdot e}_z)$, their dependence on $Pr$ can differ because the skewness of $B_z$ is influenced by the magnitudes of $B_z$ whereas the alignments $\bm{e_B}\bm{\cdot e}_z$ are not.

For the passive scalars which are in the quasi-stationary regime, the results in figure \ref{fig:alignment} also show that $\bm{e_B}\bm{\cdot e}_z>0$ events are the most probable for $Pr=1$. However, the bias towards $\bm{e_B}\bm{\cdot e}_z>0$ events becomes much weaker in going from $Pr=1$ to $Pr=7$, and this is consistent with previous results that show that for fixed $Re$, the ramp-cliffs become weaker as $Pr$ is increased beyond one \citep{buaria20a,shete22}. It is interesting to note, however, that the results for $Pr=7$ show that $\varphi(\bm{e_B}\bm{\cdot e}_z)$, while almost uniform for $|\bm{e_B}\bm{\cdot e}_z|\lesssim 0.9$, is strongly non-uniform for $|\bm{e_B}\bm{\cdot e}_z|> 0.9$. This residual preferential alignment is likely due to extreme regions of the flow with weak fluctuating scalar gradients where $\|\bm{B}\|\leq O(\beta)$ even though $\sigma_B\gg \beta$, since in such regions the mean-scalar gradient would still influence $\bm{B}$. However, the probability of such regions becomes vanishingly small for $\sigma_B/\beta\to\infty$, in which limit we would expect a uniform PDF $\varphi(\bm{e_B}\bm{\cdot e}_z)$.

Further insights into the role of buoyancy on the velocity gradient dynamics can be obtained by considering the relative importance of the nonlinear amplification and buoyancy terms in regions classified by the invariant $Q\equiv -\bm{A\cdot A}/2$. Regions where $Q>0$ are rotation (or vorticity) dominated regions, while $Q<0$ are strain dominated regions. The contributions to $\langle \mathcal{P}_{A1}\rangle$ and $-Fr^{-2}\langle \mathcal{P}_{B2}\rangle_Q$ from different regions may be considered using the decompositions
\begin{align}
\langle\mathcal{P}_{A1}\rangle &=\int_\mathbb{R}\varphi(Q)\langle\mathcal{P}_{A1}\rangle_Q\,dQ,\\
\langle\mathcal{P}_{B2}\rangle &=\int_\mathbb{R}\varphi(Q)\langle\mathcal{P}_{B2}\rangle_Q\,dQ,
\end{align}
where $\varphi(Q)$ is the PDF of $Q$. In a neutral flow, $\langle\mathcal{P}_{A1}\rangle_Q$ would be positive for $Q>0$ because of the prevalence of vortex stretching over vortex compression, and for $Q<0$, $\langle\mathcal{P}_{A1}\rangle_Q$ should also be positive but now because of the prevalence of strain self-amplification over against suppression, which is associated with the intermediate eigenvalue of the strain-rate tensor being positive on average \citep{tsinober_book,tsinober01}. On the other hand, while the integral of $\varphi(Q)\langle\mathcal{P}_{B2}\rangle_Q$ over all $Q$ is negative, there is no reason why $\varphi(Q)\langle\mathcal{P}_{B2}\rangle_Q$ must be negative for all $Q$. If it does not, then this would mean that buoyancy can have opposite effects on velocity gradient amplification in strain and rotation dominated regions of the flow. 

\begin{figure}
\includegraphics{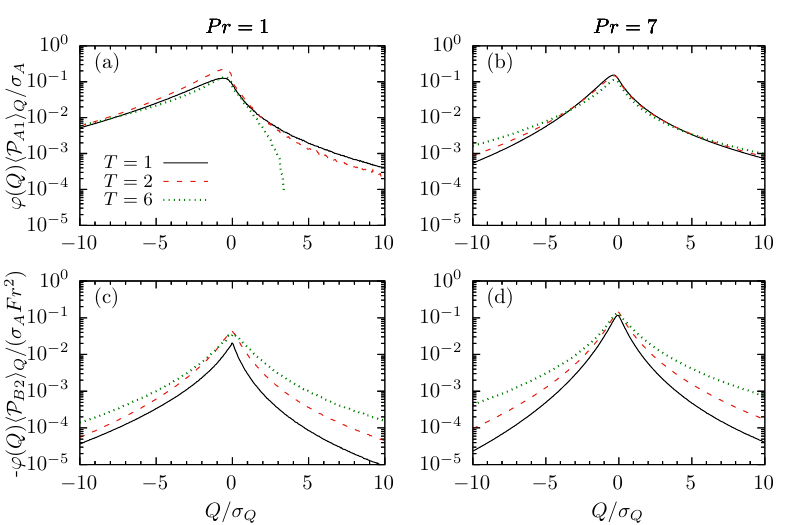}
\caption{Results for (a), (b) $\sigma_A^{-1}\varphi(Q)\langle\mathcal{P}_{A1}\rangle_Q$ and (c), (d) $-\sigma_A^{-1} Fr^{-2}\varphi(Q)\langle\mathcal{P}_{B2}\rangle_Q$ from stratified DNS. Plots (a),(c) are for $Pr=1$, plots (b),(d) are for $Pr=7$, and different curves are for different buoyancy times $T$. Note that for $Pr=1$, $\sigma_A^{-1}\varphi(Q)\langle\mathcal{P}_{A1}\rangle_Q$ becomes negative at $T=6$ for $Q/\sigma_Q\gtrsim 2$.
\label{fig:sstCavg}}
\end{figure}

In figure \ref{fig:sstCavg} we plot $\sigma_A^{-1}\varphi(Q)\langle\mathcal{P}_{A1}\rangle_Q$ and $-\sigma_A^{-1} Fr^{-2} \varphi(Q)\langle\mathcal{P}_{B2}\rangle_Q$, whose integrals over all $Q$ yield $\langle\mathcal{P}_{A1}\rangle$ and $-Fr^{-2}\langle\mathcal{P}_{B2}\rangle$, respectively. Consistent with the behavior in neutral flows, the results imply that in most cases $\langle\mathcal{P}_{A1}\rangle_Q$ is positive for all $Q$, and so in both strain and vorticty dominates regions of stratified turbulence, the average effect of $\mathcal{P}_{A1}$ is to amplify the velocity gradients. However, for $Pr=1$ and $Q>0$, the quantity $\sigma_A^{-1}\varphi(Q)\langle\mathcal{P}_{A1}\rangle_Q$ decreases significantly with increasing $T$, and at $T=6$ it becomes negative for $Q/\sigma_Q\gtrsim 2$. This implies that in regions where the vorticity is largest, vortex compression is dominating over vortex stretching, and this is why $\sigma_A^{-1}\varphi(Q)\langle\mathcal{P}_{A1}\rangle_Q$ steadily reduces for $Q>0$ as time advances. By contrast, for the $Pr=7$ case, $\sigma_A^{-1}\varphi(Q)\langle\mathcal{P}_{A1}\rangle_Q$ is almost independent of time for $Q>0$. 

For $Pr=1$, the values of $\sigma_A^{-1}\varphi(Q)\langle\mathcal{P}_{A1}\rangle_Q$ are significantly larger for $Q<0$ than for $Q>0$, and this is associated with velocity gradient production being stronger in strain dominated regions that in vorticity dominated regions (which is in turn the reason why strain self-amplification makes a larger contribution than vortex stretching to the kinetic energy cascade \citep{carbone_bragg_2020,johnson20,johnson_2021}, which is also the case in stratified turbulence \citep{zhang22}). For $Pr=7$, where the effects of buoyancy on the velocity gradient dynamics are stronger than for $Pr=1$, we see that $\sigma_A^{-1}\varphi(Q)\langle\mathcal{P}_{A1}\rangle_Q$ is much more symmetric with respect to $Q$. Compared to the $Pr=1$ case, velocity gradient production in strain dominated regions is much weaker, and that in vorticity dominated regions is much stronger for $Pr=7$.

The results for $-\sigma_A^{-1} Fr^{-2} \varphi(Q)\langle\mathcal{P}_{B2}\rangle_Q$ reveal that $\langle\mathcal{P}_{B2}\rangle_Q$ is in fact positive for all $Q$, meaning that buoyancy acts as a source for velocity gradients in both strain and vorticity dominated regions of the flow. Comparing $-\sigma_A^{-1} Fr^{-2} \varphi(Q)\langle\mathcal{P}_{B2}\rangle_Q$ for $Pr=1$ and $Pr=7$ shows that the function increases significantly at almost all $Q$ as $Pr$ is increased, just as was shown to occur for the mean value $-Fr^{-2}\langle\mathcal{P}_{B2}\rangle$ in figure \ref{fig:meanP_A}(a). Therefore, increasing $Pr$ causes the buoyancy production term to grow not only in regions of relatively low $Q/\sigma_Q$ (which dominate $-Fr^{-2}\langle\mathcal{P}_{B2}\rangle$), but also in regions of large fluctuations where $|Q/\sigma_Q|\gg 1$. In figure \ref{fig:meanP_A}(a) it was shown that for $Pr=7$, $\langle\mathcal{P}_{B2}\rangle$ and $-Fr^{-2}\langle\mathcal{P}_{B2}\rangle$ are of the same order for $T\gtrsim 1$, and almost equal for $T\gtrsim 4$. However, the results for $\sigma_A^{-1}\varphi(Q)\langle\mathcal{P}_{A1}\rangle_Q$ and $-\sigma_A^{-1} Fr^{-2} \varphi(Q)\langle\mathcal{P}_{B2}\rangle_Q$ show that the former is generally much larger than the latter when $|Q/\sigma_Q|\gg 1$ and $T\geq 1$. This means that during large fluctuations of the velocity gradients, the nonlinear amplification mechanism $\mathcal{P}_{A1}$ dominates over the buoyancy contribution $-Fr^{-2}\mathcal{P}_{B2}$. This is easily understood from the fact that the definition of $\mathcal{P}_{A1}$ involves $\bm{A}$ to the power of three, while $\mathcal{P}_{B2}$ involves $\bm{A}$ to the power of one, and therefore $\mathcal{P}_{A1}$ grows much more rapidly than $\mathcal{P}_{B2}$ when $Q\equiv -\bm{A\cdot A}/2$ is driven to large values.

\FloatBarrier

\begin{figure}
\includegraphics{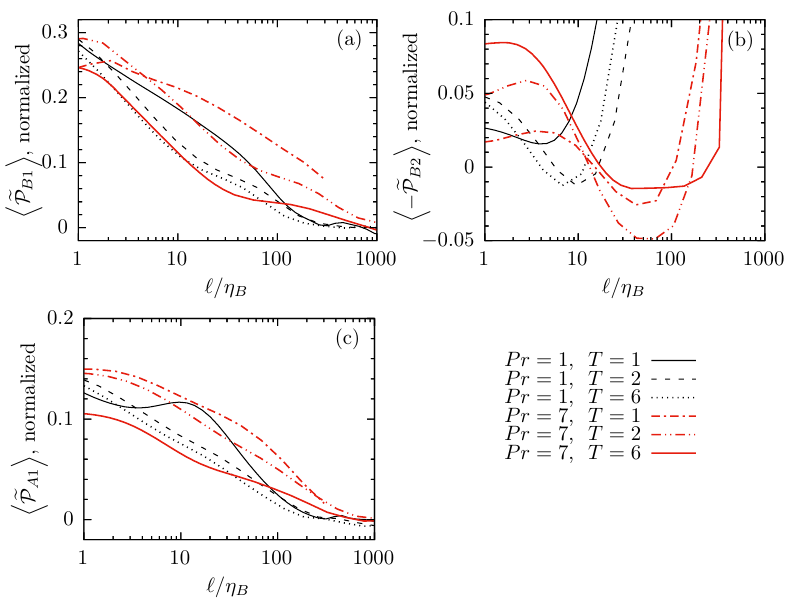}
\caption{Results for the filtered production terms (a) $\langle\widetilde{\mathcal{P}}_{B1}\rangle\equiv-\langle\widetilde{\bm{B}}\bm{\cdot}\widetilde{\bm{A}}^\top\bm{\cdot}\widetilde{\bm{B}}\rangle$, (b) $\langle\widetilde{\mathcal{P}}_{B2}\rangle\equiv\beta\langle\widetilde{\bm{B}}\bm{\cdot}\widetilde{\bm{A}}^\top\bm{\cdot}\bm{e}_z\rangle$, (c) $\langle\widetilde{\mathcal{P}}_{A1}\rangle\equiv-\langle\widetilde{\bm{A}}^\top\bm{:}(\widetilde{\bm{A}}\bm{\cdot}\widetilde{\bm{A}})\rangle$. Results for the first two quantities are normalized using $\sigma_{\widetilde{A}}\sigma_{\widetilde{B}}^2$ while the third is normalized using $\sigma_{\widetilde{A}}^3$, where $\sigma_{\widetilde{A}}\equiv\sqrt{\langle\|\widetilde{\bm{A}}\|^2\rangle}$ and $\sigma_{\widetilde{B}}\equiv\sqrt{\langle\|\widetilde{\bm{B}}\|^2\rangle}$.
\label{fig:filt}}
\end{figure}

Finally, in \S\ref{TSST} we argued that the filtered buoyancy production term $-Fr^{-2}\langle\widetilde{\mathcal{P}}_{B2}\rangle\equiv-Fr^{-2}\beta\langle\widetilde{\bm{B}}\bm{\cdot}\widetilde{\bm{A}}^\top\bm{\cdot}\bm{e}_z\rangle$ will change sign as the filter length $\ell$ increases in a stationary flow. In particular, for $\lim_{\ell/\eta_B\to0}\langle\widetilde{\mathcal{P}}_{B2}\rangle\to \langle{\mathcal{P}}_{B2}\rangle$ which is negative, but when $\ell/\eta_B$ becomes large enough for $|\langle\widetilde{\mathcal{P}}_{B1}\rangle|\ll |\langle\widetilde{\mathcal{P}}_{B2}\rangle|$ then $\langle\widetilde{\mathcal{P}}_{B2}\rangle$ must become positive because in this range it must act as the dominant source term in the equation for $\langle\|\widetilde{\bm{B}}\|^2\rangle$ in the stationary regime. The implication of this is that in the equation for $\langle\|\widetilde{\bm{A}}\|^2\rangle$, the buoyancy term $-Fr^{-2}\langle\widetilde{\mathcal{P}}_{B2}\rangle$ acts as a source term at sufficiently small $\ell/\eta_B$, while it acts as a sink term at large $\ell/\eta_B$. To test this, in figure \ref{fig:filt} we plot $\langle\widetilde{\mathcal{P}}_{B1}\rangle\equiv-\langle\widetilde{\bm{B}}\bm{\cdot}\widetilde{\bm{A}}^\top\bm{\cdot}\widetilde{\bm{B}}\rangle$, $\langle\widetilde{\mathcal{P}}_{B2}\rangle\equiv\beta\langle\widetilde{\bm{B}}\bm{\cdot}\widetilde{\bm{A}}^\top\bm{\cdot}\bm{e}_z\rangle$, and $\langle\widetilde{\mathcal{P}}_{A1}\rangle\equiv-\langle\widetilde{\bm{A}}^\top\bm{:}(\widetilde{\bm{A}}\bm{\cdot}\widetilde{\bm{A}})\rangle$, suitably normalized using $\sigma_{\widetilde{A}}\equiv\sqrt{\langle\|\widetilde{\bm{A}}\|^2\rangle}$ and $\sigma_{\widetilde{B}}\equiv\sqrt{\langle\|\widetilde{\bm{B}}\|^2\rangle}$.

The results show that $\langle\widetilde{\mathcal{P}}_{B1}\rangle$ and $\langle\widetilde{\mathcal{P}}_{A1}\rangle$ are positive at all scales in the flow, and so act as source terms at all scales in the equations for $\langle\|\widetilde{\bm{B}}\|^2\rangle$ and $\langle\|\widetilde{\bm{A}}\|^2\rangle$, respectively. The main difference between the stratified results and the passive scalar results for $\langle\widetilde{\mathcal{P}}_{B1}\rangle$ and $\langle\widetilde{\mathcal{P}}_{A1}\rangle$ is that for the passive case these quantities  (when normalized as in the plot) do not significantly reduce until much larger values of $\ell/\eta_B$. This is mainly due to the flow Reynolds number, and hence $L/\eta_B$, being much larger for the passive scalar runs. The results for $\langle\widetilde{\mathcal{P}}_{B2}\rangle$ for the stratified DNS show that this term changes sign as $\ell/\eta_B$ is increased, such that the buoyancy term $-Fr^{-2}\langle\widetilde{\mathcal{P}}_{B2}\rangle$ acts as a source term for $\langle\|\widetilde{\bm{A}}\|^2\rangle$ at small-scales, but as a sink term at larger scales. Although this agrees with the prediction from \S\ref{TSST}, the conditions under which the sign change is observed to occur disagrees with those predicted by the analysis. In particular, although $\langle\widetilde{\mathcal{P}}_{B2}\rangle$ becomes positive as $\ell/\eta_B$ increases, it becomes negative again at even larger $\ell/\eta_B$, even though $|\langle\widetilde{\mathcal{P}}_{B1}\rangle|\ll |\langle\widetilde{\mathcal{P}}_{B2}\rangle|$ at these larger scales. This disagreement is, however, almost certainly due to the fact that the analysis in \S\ref{TSST} applies to a stationary flow, whereas the DNS for stratified flow is decaying. As a result, in view of the analysis in \S\ref{Effect_Re_Pr}, $\langle\widetilde{\mathcal{P}}_{B2}\rangle$ need not be positive at scales where $|\langle\widetilde{\mathcal{P}}_{B1}\rangle|\ll |\langle\widetilde{\mathcal{P}}_{B2}\rangle|$ in order to balance $\langle \widetilde{\bm{B}}\bm{\cdot}\bm{\nabla\nabla\cdot}\bm{\tau}_\phi\rangle$ because of the contribution from $\partial_t\langle\|\widetilde{\bm{B}}\|^2\rangle<0$ at larger scales in the decaying flow.

For the passive scalar cases (not shown), $\langle\widetilde{\mathcal{P}}_{B2}\rangle$ remains negative at all scales, which is contrary to expectation based on the analysis in \S\ref{Effect_Re_Pr}. The most likely reason for this discrepancy is that since $\lim_{\ell/\eta_B \to0}\beta/\sqrt{\langle\|\widetilde{\bm{B}}\|^2\rangle}=\beta/\sqrt{\langle\|{\bm{B}}\|^2\rangle}$ is very small for the passive scalar cases, then the condition under which $\langle\widetilde{\mathcal{P}}_{B2}\rangle$ is predicted to become positive, namely $\beta/\sqrt{\langle\|\widetilde{\bm{B}}\|^2\rangle}\geq O(1)$, may only occur at $\ell=O(L)$. At such filter scales, the data for $\langle\widetilde{\mathcal{P}}_{B2}\rangle\equiv\beta\langle\widetilde{\bm{B}}\bm{\cdot}\widetilde{\bm{A}}^\top\bm{\cdot}\bm{e}_z\rangle$ will be strongly affected by statistical noise due to the box size because although theoretically $\lim_{\ell/L \to \infty}\widetilde{\bm{B}}\to\bm{0}$ and $\lim_{\ell/L \to \infty}\widetilde{\bm{A}}\to\bm{0}$ for a homogeneous flow, in practice these limiting behaviours may be approximately satisfied for $\ell\geq O(L)$. A much larger domain may therefore be required to observe $\langle\widetilde{\mathcal{P}}_{B2}\rangle$ becoming positive for the passive scalar case in order to minimize the effects of statistical noise at $\ell=O(L)$, as well as to more fully satisfy the assumptions made in the theoretical analysis of a statistically stationary, homogeneous flow.

\FloatBarrier

\section{Conclusions}

This study was primarily motivated by recent direct numerical simulations (DNS) of stably stratified turbulence that showed that as $Pr$ is increased from 1 to 7, the mean turbulent potential energy dissipation rate $Fr^{-2}\langle\chi\rangle$ (where $Fr$ is the Froude number) drops dramatically, while the mean turbulent kinetic energy dissipation rate $\langle\epsilon\rangle$ increases significantly \citep{riley23}. To understand the mechanism responsible for this surprising behavior, we analyzed the equations governing the fluctuating velocity gradient $\bm{A}$ and fluctuating density gradient $\bm{B}$. This was done for both passive scalars driven by a mean scalar gradient and stably stratified flows in order to understand the extent to which the behavior observed for stratified flows is simply due to the effects of an imposed mean scalar gradient versus the particular dynamical effects due to buoyancy forces. The predictions from the analysis were then compared with DNS results for passive scalars and stably stratified turbulence.

Production mechanisms in the equation for $\|\bm{B}\|^2$ (whose average is proportional to the mean scalar dissipation rate $\langle\chi\rangle$) are associated with the stirring processes that intensify flow gradients, and the magnitude of the resulting gradients determines the mixing rates. Prandtl number effects on the mixing rates can therefore be understood at a fundamental level by examining the effects of $Pr$ on the production mechanisms, of which there are two; one associated with $\bm{B}$, which we refer to as $\mathcal{P}_{B1}$, and the other associated with the mean scalar gradient $\beta$, which we refer to as $\mathcal{P}_{B2}$. In the passive scalar context, we discussed that $\mathcal{P}_{B1}$ is affected by a de-localization effect due to a disparity between the smallest scales of the velocity and scalar fields when $Pr\neq 1$. This de-localization effect renders $\mathcal{P}_{B1}$ less effective in amplifying $\|\bm{B}\|^2$ as $Pr$ is increased. We also argued that on average $\mathcal{P}_{B2}$ actually opposes the amplification of $\|\bm{B}\|^2$, and that this is due to the effect of the ramp-cliff structures in the scalar field. The impact of this production term depends upon the parameter regime of the flow, but when it is important, its oppositional effect causes $\langle\chi\rangle$ to decrease with increasing $Pr$. Our DNS results for $Re_\lambda=633$ and $Pr\in[0.1,7]$ show that on average $\mathcal{P}_{B2}$ does indeed oppose the production of $\|\bm{B}\|^2$, however, its contribution is negligible compared with $\mathcal{P}_{B1}$. A weak dependence of $\langle\chi\rangle$ on $Pr$ was observed which is mainly due to the de-localization effect.

For stably stratified flows where the scalar field is the fluid density, the buoyancy term in the equation for $\|\bm{A}\|^2$ is $-Fr^{-2}\mathcal{P}_{B2}$. Since on average $\mathcal{P}_{B2}$ is negative, then the effect of buoyancy is to amplify $\|\bm{A}\|^2$ on average. This is surprising because in stably stratified flows, buoyancy is expected to suppress turbulent motion. However, by analyzing the filtered velocity gradient equation we demonstrated that while buoyancy amplifies the small-scale velocity gradients, it suppresses the large-scale velocity gradients. This analysis was confirmed (???) using DNS, and is also connected with the observation in \citet{legaspi20} based on numerical simulations that there is a transfer of potential to kinetic energy at the smallest scales in stably stratified turbulence which depends on $Pr$.

Concerning the effect of $Pr$ on $\langle\epsilon\rangle$ and $\langle\chi\rangle$ in stratified turbulence, we presented an analysis for the weak-coupling regime where the effects of buoyancy on $\|\bm{A}\|^2$ and $\|\bm{B}\|^2$ are perturbative. This analysis predicts that as $Pr$ is increased, the buoyancy term in the equation for $\|\bm{A}\|^2$ should grow in strength, with the result that $\langle\epsilon\rangle$ should increase and $\langle\chi\rangle$ should decrease with increasing $Pr$, in qualitative agreement with the results in \citet{riley23}. Guided by the results and insights from the analysis, we used DNS data of stably stratified turbulence with $Pr=1$ and $Pr=7$ (the same data set used in \citet{riley23}) to compute the production terms in the equations for $\|\bm{A}\|^2$ and $\|\bm{B}\|^2$ to see how they are impacted by $Pr$ and how they differ from the passive scalar case. For $\|\bm{B}\|^2$, the results show that $\mathcal{P}_{B2}$ plays a much larger role in stratified flows than for passive scalars. This is the main reason why $\langle\chi\rangle$ is much more strongly dependent on $Pr$ in stratified flows than neutral flows, and the fact that this term opposes the production of $\|\bm{B}\|^2$ on average is the reason why $\langle\chi\rangle$ decreases with increasing $Pr$. For $\|\bm{A}\|^2$, the DNS results show that the buoyancy term $-Fr^{-2}\mathcal{P}_{B2}$ increases significantly with increasing $Pr$, in qualitative agreement with the weak-coupling analysis. This growth of the buoyancy term is the reason why $\langle\epsilon\rangle$ increases with increasing $Pr$.

We also argued that the strong effect of $Pr$ in stratified flows means that the buoyancy Reynolds number $Re_b$ and the activity parameter $Gn$ may not provide a reliable way to predict the impact of buoyancy on the smallest-scales of stably stratified turbulence. By analyzing the equation for $\bm{A}$, we proposed a new non-dimensional number $\mathcal{R}_b$ that compares the buoyancy and inertial terms in this equation and captures the effect of $Pr$. Using DNS data we showed that $\mathcal{R}_b$ provides a more reliable way to gauge whether the effects of buoyancy at the smallest scales of a stratified flow are important. Indeed, while $\mathcal{R}_b$ correctly predicts that when $Pr$ increases, the effects of buoyancy at the smallest scales increase, $Gn$ incorrectly predicts the opposite. 

Finally, an analysis of the filtered gradient equations predicted that the mean density gradient term must change sign at sufficiently large scales, such that buoyancy will act as a source for velocity gradients at small scales, but as a sink at large scales. Our DNS confirmed that there is indeed a range of scales where this buoyancy term becomes negative, however, the conditions under which this is observed to occur does not agree with those predicted by the theoretical analysis. We argued that this is most likely because while the analysis assumes a statistically stationary flow, the DNS is for decaying stratified turbulence. At larger scales where the decay term is significant in the filtered gradient equations, this changes the dominant balance of the equations relative to the stationary case, and therefore the scales at which the buoyancy term will change sign.

The analysis suggests that in the limit $\mathcal{R}_b\to\infty$, the velocity and density gradient fields in stratified turbulent flows will behave like those for a neutral flow where density is passive. In this regime, $\langle\epsilon\rangle$ will become independent of $Pr$, as will $\langle\chi\rangle$ if the large-scale Reynolds number of the flow $Re$ is also sufficiently high. However, DNS at higher $Re$ and $Pr$ are needed in order to understand how quickly this asymptotic regime is attained, and therefore whether  $\langle\epsilon\rangle$ and $\langle\chi\rangle$ might become independent of $Pr$ in parameter regimes relevant to real stratified flows. Another important topic to be explored in future work is how the results and insights from this work that focuses on the gradient field dynamics connects to the multiscale behavior of the kinetic and potential energy fields in stratified flows. In particular, does the positive contribution of buoyancy to the production of fluctuating velocity gradients imply that at the smallest scales potential energy is transferred back to the kinetic energy field, and if so, over what scales does this occur and how does it depend on $Pr$?

\backsection[Acknowledgements]{This research used resources of the Oak Ridge Leadership Computing Facility at the Oak Ridge National Laboratory, which is supported by the Office of Science of the U.S. Department of Energy under Contract No. DE-AC05-00OR22725.  Additional resources were provided
through the U.S.\ Department of Defense High Performance Computing Modernization
Program by the Army Engineer Research and Development Center and the Army
Research Laboratory under Frontier Project FP-CFD-FY14-007.}

\backsection[Funding]{ADB was supported by National Science Foundation (NSF) CAREER award \# 2042346. SdeBK was supported by U.S. Office of Naval Research Grant number N00014-19-1-2152. 
}

\backsection[Declaration of interests]{The authors report no conflict of interest.}

\bibliographystyle{jfm}
\bibliography{bib}

\end{document}